# Decoherence in a superconducting quantum bit circuit

G. Ithier,[1] E. Collin,[1] P. Joyez,[1] P.J. Meeson,[1,2] D. Vion,[1] D. Esteve,[1] F. Chiarello,[3] A. Shnirman,[4] Y. Makhlin,[4,5] Josef Schriefl,[4,6] and G. Schön[4]

[1]*Quantronics group, Service de Physique de l'Etat Condensé,*
*DSM/DRECAM, CEA Saclay, 91191 Gif-sur-Yvette, France.*
[2]*Department of Physics, Royal Holloway, University of London,*
*Egham Hill, Egham, Surrey TW20 0EX, U. K.*
[3]*Istituto di Fotonica e Nanotecnologia, CNR,*
*Via Cineto Romano, 42 00156 Roma, Italy.*
[4]*Institut für Theoretische Festkörperphysik,*
*Universität Karlsruhe, D-76128 Karlsruhe, Germany.*
[5]*Landau Institute for Theoretical Physics,*
*Kosygin st. 2, 119334 Moscow, Russia.*
[6]*Ecole Normale Supérieure de Lyon,*
*46 allée d'Italie, 69364 Lyon cedex 07, France.*

(Dated: 2005, August the 2nd)

## Abstract

Decoherence in quantum bit circuits is presently a major limitation to their use for quantum computing purposes. We present experiments, inspired from NMR, that characterise decoherence in a particular superconducting quantum bit circuit, the quantronium. We introduce a general framework for the analysis of decoherence, based on the spectral densities of the noise sources coupled to the qubit. Analysis of our measurements within this framework indicates a simple model for the noise sources acting on the qubit. We discuss various methods to fight decoherence.

# I. INTRODUCTION

It has been demonstrated that several types of superconducting circuits based on Josephson junctions are sufficiently quantum that simple manipulations of their quantum state[1-7] can be performed. These circuits are candidates for implementing quantum bits (qubits), which are the basic building blocks of a quantum processor. The coherence time of the quantum state is an important figure of merit, being related to the number of qubit operations that can be performed without error. Despite significant advances in coherence times during recent years, with coherence times of order $0.5$ $\mu s$ reached, decoherence due to the coupling between the quantum circuit and the degrees of freedom of the environment still severely hinders using these circuits for the development of a quantum processor[8], even with a small number of qubits. Thus the quantitative characterization and understanding of decoherence processes is presently a central issue for the development of qubit circuits. In this work, we present experiments that characterize the sources of decoherence in a particular qubit circuit, the quantronium. We also develop a general framework for the theoretical analysis of such data, a framework that can be adapted to other circuits. We address the problem of decoherence both during the free evolution of the qubit and during its driven evolution when coupled to a small AC excitation. For these two situations, we also consider particular control sequences that aim at maintaining quantum coherence. The analysis of our data leads to a simple model for the spectral densities of the noise sources coupled to the qubit.

The paper is organized as follows: In Section II, the quantronium device is introduced and manipulation and readout of its quantum state are described; the experimental setup is presented, and the principal noise sources responsible for decoherence are discussed. In section III, a general framework is introduced for the description of decoherence processes in the two situations of free and driven evolutions, for both linear and quadratic coupling of the qubit to variations in the control parameters. In sections IV and V, we report experimental results on the measurement of decoherence in all these situations, and analyze them within the theoretical framework of section III. We introduce methods inspired from Nuclear Magnetic Resonance (NMR), such as spin echoes and spin locking, which probe the spectral density of the noise sources responsible for decoherence. From this we invoke constraints on the spectral density of the noise sources and develop a simple model of environmental noise. We also discuss how to improve the quantum coherence time of a qubit. Then, section VI summarizes what has been learnt with the quantronium on decoherence processes in Josephson qubits, and how to suppress decoherence.

## II. THE QUANTRONIUM CIRCUIT

### A Principles

The quantronium circuit[9], described in Fig. 1, combines a split Cooper Pair Box (CPB)[10-12] that plays the role of a qubit, and a hysteretic current biased Josephson junction for readout. It consists of a superconducting loop interrupted by two adjacent tunnel junctions with Josephson energies $E_J/2(1 \pm d)$, where $d$ is an asymmetry coefficient made as small as possible, and by the readout junction with a Josephson energy $\mathcal{E}_J \gg E_J$. The two small junctions define the superconducting island of the box, whose total capacitance to ground is $C_\Sigma$ and Cooper pair Coulomb energy $E_C = (2e)^2/2C_\Sigma$. The island is coupled to a voltage source $U$ through a gate capacitance $C_g$, and an external magnetic flux $\Phi$ can be applied to the loop. The biasing parameters are thus the reduced gate charge $N_g = C_g U/2e$, and the reduced flux $\phi = 2\pi\Phi/\Phi_0$ ($\Phi_0 = 2\pi\varphi_0 = h/2e$). The latter determines, together with the bias current $I_b$, the phase difference $\delta$ across the small junctions. Quantum mechanically, the number $\widehat{N}$ of excess Cooper pairs on the island, and the superconducting phase difference $\widehat{\gamma}$ across the readout junction form a set of degrees of freedom that fully characterize the system. Using also their respective conjugate variables $\widehat{\theta}$ and $\widehat{q}$, together with the phase relation $\widehat{\delta} = \widehat{\gamma} + \phi$, the Hamiltonian $\widehat{H}$ of the CPB and readout parts reads

$$\widehat{H} = \widehat{H}_{CPB} + \widehat{H}_r, \tag{1}$$

$$\widehat{H}_{CPB} = E_C(\widehat{N} - N_g)^2 - E_J \cos(\widehat{\delta}/2) \cos\widehat{\theta} + d\, E_J \sin(\widehat{\delta}/2) \sin\widehat{\theta}, \tag{2}$$

$$\widehat{H}_r = \mathcal{E}_C\, \widehat{q}^2 - \mathcal{E}_J \cos\widehat{\gamma} - \varphi_0 I_b\, \widehat{\gamma}. \tag{3}$$

Here $\mathcal{E}_C = (2e)^2/2\mathcal{C}_J$ and $\mathcal{C}_J$ are the Cooper pair Coulomb energy and the capacitance of the readout junction respectively. The coupling between both subsystems results from the phase constraint given above. Except at readout, when the bias current $I_b$ is extremely close to the critical current $\mathcal{I}_0 = \mathcal{E}_J/\varphi_0$ of the readout junction, a full quantum calculation using $\widehat{H}$ shows that the quantum nature of $\widehat{\gamma}$ can be ignored and that the approximation $\widehat{\gamma} \simeq \gamma \simeq \arcsin(I_b/\mathcal{I}_0)$ that neglects the contribution of the current $\langle\widehat{i}\rangle$ in the quantronium loop to the current in the readout junction is excellent. The CPB eigenstates are then determined only by $N_g$ and $\widehat{\delta} \simeq \delta = \gamma + \phi$. For a large range of $N_g$ and $\delta$, the energy spectrum of $\widehat{H}_{CPB}$ is anharmonic and its two lowest energy eigenstates $|0\rangle$ and $|1\rangle$ define a qubit with energy splitting $\hbar\omega_{01}(\delta, N_g)$. The Hamiltonian $\widehat{H}_{qb}$ of this qubit, i.e. the restriction of $\widehat{H}_{CPB}$ to the manifold $\{|0\rangle, |1\rangle\}$, is that of a fictitious spin 1/2 particle $\vec{\widehat{\sigma}} = \hat{\sigma}_x\vec{x} + \hat{\sigma}_y\vec{y} + \hat{\sigma}_z\vec{z}$ in an effective magnetic field $\vec{H}_0$,

$$\widehat{H}_{qb} = -\frac{1}{2}\vec{H}_0\,\vec{\widehat{\sigma}}, \tag{4}$$

in an eigenbasis that depends on the working point $(\delta, N_g)$ $\vec{H}_0 = \hbar\omega_{01}\vec{z}$. At the point $P_0 = (\delta = 0, N_g = 1/2)$, $\omega_{01}$ is stationary with respect to small variations of all the control parameters (see Fig. 1), which makes the quantronium almost immune to decoherence, as previously demonstrated[2,3]. $P_0$ is therefore an optimal point for manipulating the quantronium state in a coherent way.

## 1 Manipulation of the quantum state

The manipulation of the quantronium state is achieved by varying the control parameters $N_g$ and/or $I_b$, either in a resonant way at a microwave angular frequency $\omega_{\mu w}$ close to the transition frequency $\omega_{01}$, or adiabatically. In the resonant scheme, a microwave pulse is applied to the gate and induces the variation $\Delta N_g \cos(\omega_{\mu w} t + \chi)$, where $\chi$ is the phase of the microwave with respect to a reference carrier. The qubit dynamics is conveniently described using the Bloch sphere in a frame rotating at $\omega_{\mu w}$ (see Fig. 1), where the effective magnetic field becomes $\vec{H}_0 = \hbar \Delta \omega \, \vec{z} + \hbar \omega_{R0} \left[ \vec{x} \cos \chi + \vec{y} \sin \chi \right]$, with $\Delta \omega = \omega_{01} - \omega_{\mu w}$ being the detuning and $\omega_{R0} = 2 E_C \Delta N_g \left| \langle 1 | \, \widehat{N} \, | 0 \rangle \right| / \hbar$ the Rabi frequency. At $\Delta \omega = 0$, pure Rabi precession takes place around an axis lying in the equatorial plane and making an angle $\chi$ with respect to the $X$ axis. Then, any single-qubit operation can be performed by combining three rotations around the $X$ and $Y$ axes[8,13]. The sequences used to characterize decoherence in this work involve principally two types of pulses, namely $\pi/2$ and $\pi$ rotations around the $X$ or $Y$ axes. Between microwave pulses, the free evolution of the spin corresponds to a rotation around the $Z$ axis at frequency $-\Delta \omega$. Such $Z$ rotations can also be induced with the adiabatic method[13], that is by adiabatically varying the transition frequency. This can be achieved by applying a pulse that satisfies the adiabaticity criterion $\left| d\lambda/dt \, \langle 1 | \, \partial \hat{H}_{qb} / \partial \lambda \, | 0 \rangle \right| / (\hbar \omega_{01})[14] \ll \omega_{01}$ to one of the reduced parameters $\lambda = N_g$ or $\lambda = \delta/2\pi$.

## 2 Readout

For readout, the quantronium state is projected onto the $|0\rangle$ and $|1\rangle$ states, which are then discriminated through the difference in their supercurrents $\left\langle \hat{i} \right\rangle$ in the loop[2]. The readout junction is actually used to transfer adiabatically the information about the quantum state of the qubit onto the phase $\gamma$, in analogy with the Stern & Gerlach experiment, in which the spin state of a silver atom is entangled with its transverse position. For this transfer, a trapezoidal bias current pulse $I_b(t)$ with a maximum value $I_M$ slightly below $\mathcal{I}_0$ is applied to the circuit. Starting from $\delta = 0$, the phases $\gamma$ and $\delta$ grow during the current pulse and the state-dependent supercurrent $\left\langle \hat{i} \right\rangle$ develops in the loop. This current adds algebraically to $I_b$ in the large junction and thus modifies its switching rate $\Gamma$. By precisely adjusting $I_M$ and the duration of the pulse, the large junction switches during the pulse to a finite voltage state with a large probability $p_1$ for state $|1\rangle$ and with a small probability $p_0$ for state $|0\rangle$[9]. A switching or non switching event is detected by measuring the voltage across the readout junction with a room temperature amplifier, and the switching probability $p$ is determined by repeating the experiment. Note that at the temperature used in this work, and for the readout junction parameters, switching occurs by Macroscopic Quantum Tunneling (MQT) of the phase $\gamma$[15,16]. The theoretical error rate in discriminating the two qubit states is expected to be lower than 5 % at temperature $T \leq 40$ mK for the parameters of our experiment[9]. Note than the present readout scheme does not implement a quantum non demolition (QND) measurement since the quantronium quantum states are fully destroyed when the readout junction switches. An alternative quantronium readout designed to be QND has been developed[17] after this work.

## B   Experimental implementation

The quantronium sample used for this work was fabricated using standard e-beam lithography and double angle shadow evaporation of aluminum. Scanning electron micrographs of its whole loop (with area $\sim 6~\mu\mathrm{m}^2$) and of the island region are shown in Fig. 2 with the schematic experimental setup. The quantronium loop was deposited on top of four gold pads designed to trap spurious quasiparticles in the superconductor, including those generated by the switching of the readout junction. This junction was also connected in parallel to an on-chip interdigitated gold capacitor $\mathcal{C}_J \simeq 0.6$ pF, designed to lower its bare plasma frequency to approximately 7 or 8 GHz. Separate gates with capacitances 40 and 80 aF were used for the DC and microwave $N_g$ signals, respectively. The sample was mounted in a copper shielding box thermally anchored to the mixing chamber of a dilution refrigerator with base temperature 15 mK. The impedance of the microwave gate line as seen from the qubit was defined by a 50 $\Omega$ attenuator placed at 600 mK. That of the DC gate line was defined below 100 MHz by a 1 k$\Omega$ resistor at 4 K, and its real part was measured to be close to 80 $\Omega$ in the $6 - 17$ GHz range explored by the qubit frequency. The bias resistor of the readout junction, $R_b = 4.1$ k$\Omega$, was placed at the lowest temperature. Both the current biasing line and the voltage measurement lines were shunted above a few 100 MHz by two surface mounted 150 $\Omega$ − 47 pF RC shunts located a few millimeters away from the chip. These shunts define the quality factor $Q$ of the readout junction. The external magnetic flux $\Phi$ was produced by a superconducting coil with a self-inductance $L = 0.12$ H, placed 3 mm from the chip, and whose mutual inductance with the quantronium loop was $M = 0.14$ pH. To filter current noise in this coil, a 50 $\Omega$ shunt resistor was placed at 1 K. The sample holder and its coil were magnetically shielded by a 3 mm-thick superconducting aluminum cylinder open at one end and supported by a screw from the sample-holder. The whole assembly is placed in a second copper box also attached to the mixing chamber.

The microwave gate pulses used to manipulate the qubit were generated by mixing continuous microwaves with 1 ns rise time trapezoidal pulses with variable duration $\tau$, defined here as the width at half maximum. With the 60 dB attenuation of the microwave gate line, the range of Rabi frequencies $\omega_{R0}$ that was explored extends up to 250 MHz. The switching probability $p$ was averaged over 25 000 − 60 000 events, chosen to obtain good statistics, with a repetition rate in the 10 − 60 kHz range, slow enough to allow quasiparticle retrapping. The electronic temperature during operation, $T_e \sim 40$ mK, and the relevant parameters $E_J = 0.87$ $k_B$ K, $E_C = 0.66$ $k_B$ K, $\omega_{01}(P_0)/2\pi = 16.41$ GHz, $d \sim 3 - 4$ %, $\mathcal{I}_0 = 427$ nA, and $Q \sim 3$ were measured as reported in a previous work[16] by fitting spectroscopic data, such as that shown in Fig. 11. Figure 3 shows a typical gate pulse, Rabi oscillations of the switching probability $p$, and a check of the proportionality between the Rabi frequency $\omega_{R0}$ and $\Delta N_g$. The loop currents $i_0$ and $i_1$ of the two qubit states, calculated using these values of $E_J$, $E_C$, and $d$ are shown in Figure 4.

The readout was performed with 100 ns wide pulses (Fig. 4) giving a switching probability in the 10 % − 90 % range at $\gamma_M \sim 72$ °. At the top of the readout pulse, the phase $\delta$ was close to $\delta_M \simeq 0.37 \times 2\pi \simeq 130$ °, where the difference between the loop currents for the two qubit states is the largest, and where the sensitivity was experimentally maximal. To reach this $\delta_M$ value starting from any value $\delta_{op}$ where the qubit was operated, we had to set $I_b$ and $\Phi$ at the values $I_{op}$ and $\Phi_{op}$ such that $\delta_{op} = \Phi_{op} + \gamma_{op}$ and $\delta_M = \Phi_{op} + \gamma_M$. The corresponding bias current $I_b$ pulse when the quantronium is operated at the optimal point $P_0$ is shown in the top panels of Fig. 4: A negative 'pre-bias' current $I_{op}$ is used to compensate a positive flux. The fidelity $\eta$ of the measurement, i.e. the largest value of $p_1 - p_0$, was $\eta \approx 0.3 - 0.4$ in the present series of experiments, as shown in the bottom-right panel of Figure 4. Although $\eta$ is larger than in our previous work[2], this fidelity is nevertheless much smaller than the 0.95 expected. This loss, also observed in other Josephson qubits e.g.[5], is attributed to spurious relaxation during the adiabatic ramp used to switch the readout on. Indeed, the signal loss after three adjacent short microwave $\pi$ pulses is approximately the same as after one. Moreover, it was found in some Josephson qubit experiments[5,18] that the fidelity is improved when increasing the readout speed. The shape of the $p(I_M)$ curve after a $\pi$ pulse (see Fig. 4) also shows that the fidelity loss increases with $p$, which leads to a slight asymmetry of $p$ oscillations in most of the experiments presented here (see for instance the lack of signal at the top of the oscillations on Figs. 8 and 16). This asymmetry limits the accuracy of our decoherence rate measurements. In order to minimize its effect, we have chosen to use the bottom of the envelopes of the $p$ oscillations to quantify decoherence.

## C   Decoherence sources

Like any other quantum object, the quantronium qubit is subject to decoherence due to its interaction with uncontrolled degrees of freedom in its environment, including those in the device itself. These degrees of freedom appear as noise induced in the parameters entering the qubit Hamiltonian (2), i.e. the Josephson energy $E_J$, the gate charge $N_g$, or the superconducting phase difference $\delta$. Using dimensionless parameters $\lambda = E_J/E_{J0}$ ($E_{J0}$ being the nominal $E_J$), $\lambda = N_g$, or $\lambda = \delta/(2\pi)$, each noise source is conveniently described by its quantum spectral density $S_\lambda(\omega) \equiv 1/(2\pi) \int dt \langle \widehat{\delta\lambda}(0)\widehat{\delta\lambda}(t) \rangle e^{-i\omega t}$, where $\widehat{\delta\lambda}$ is regarded here as an operator acting on environmental variables. This function quantifies the ability of the source to absorb and to emit an energy quantum $\hbar|\omega|$, at positive and negative $\omega$, respectively. The symmetrized spectral density $Ss_\lambda(\omega) = 1/2[S_\lambda(-\omega) + S_\lambda(\omega)]$ and its classical limit $Sc_\lambda(\omega)$ at $k_B T \gg \hbar\omega$ will be also used. Decoherence of the qubit will be described here in terms of energy exchange with a noise source on one hand, and in terms of random dephasing between states $|0\rangle$ and $|1\rangle$ due to adiabatic variations of the transition frequency, on the other hand. Taking into account that $k_B T \ll \hbar\omega_{01}$ in our experiment, we distinguish relaxation processes involving $S_\lambda(+\omega_{01})$ and dephasing processes involving $Sc_\lambda(|\omega| \ll \omega_{01})$. The main noise sources acting in the quantronium are schematically depicted in Fig. 5, and their spectral densities are discussed below. Since 'pure' dephasing (see section III) dominates decoherence, special attention is paid to the low frequency part $Sc_\lambda(\omega \approx 0)$.

### 1  Noise in $E_J$: Two level fluctuators in the tunnel barriers

A first source of decoherence arises from the fluctuations of the Josephson energy $E_J$ of the two small junctions. The associated critical current noise, which has not been measured in our sample, has been characterized, at frequencies $f$ up to 10 kHz and at temperatures $T$ between 100 mK and 4 K, in various Josephson devices[19,20] made of different materials and with different technologies, including those used in this work. The Josephson energy noise is empirically described, for a single junction with critical current $\mathcal{I}_0$, by a $1/f$ spectral density that scales with $\mathcal{I}_0^2$, $T^2$, and with the inverse of the junction area. Extrapolating the results reported in Refs.[19,21] to the maximum electronic temperature $T_e$ of the quantronium during its operation, i.e. 40 mK, leads to an estimate for the spectral density of relative $E_J$ fluctuations: $Sc_{\delta E_J/E_J}(|\omega| < 2\pi \times 10 \text{ kHz}) \sim (0.5 \times 10^{-6})^2/|\omega|$. The critical current noise is presently attributed to atomic defects located in the oxide of the tunnel junctions. A simple model assumes that these defects are Two-Level Fluctuators (TLF) switching between two states that correspond to an open and to a closed tunneling channel through the junction. The distribution in the energy splitting of these TLFs is thought to be very broad and to extend above the transition energies of Josephson qubits. This picture is supported by the observation of a coherent coupling between a phase qubit and uncontrolled TLFs randomly distributed in frequency[22]. With the quantronium sample used in this work, the authors have also observed in one of the experimental runs an avoided level crossing in the spectroscopic data, which demonstrated a strong coupling between the qubit and an unknown TLF that was later eliminated by annealing the sample at room temperature. These observations suggest that TLFs located in the tunnel barriers not only generate low frequency $E_J$ noise, but can also play an important role in the relaxation of Josephson qubits.

### 2 Noise in $N_g$: Background charged two-level fluctuators and gate line impedance

A second source of decoherence is the noise on the gate charge $N_g$. Like any Coulomb blockade device, the quantronium is subject to Background Charge Noise (BCN) due to microscopic charged TLFs acting as uncontrolled additional $N_g$ sources. Although the whole collection of TLFs produces a noise whose spectral density approximately follows a $1/f$ law[23-25], telegraph noise due to some well coupled TLF can be observed as well[10]. These well coupled TLFs are, for instance, responsible for the substructure of the quantronium resonance line recorded at $N_g \neq 1/2$ (see Fig. 11). Complementary works[26] have shown that the charged TLFs are partly located in the substrate, partly in the oxide layer covering all the electrodes, and partly in the oxide barriers of the tunnel junctions themselves. It has been suggested that some TLFs contribute both to the critical current noise and to the charge noise[27]. The typical amplitude $A$ of the spectral density $S_{CN_g}^{BCN}(|\omega| < 2\pi \times 100\,\text{kHz}) = A/|\omega|$ depends on temperature, on junction size and on the screening of the island by the other electrodes. Its value is commonly found in the range $[10^{-6}, 10^{-7}]$ for the parameters of our experiment. The amount and the energy splitting distribution of charged TLFs in Josephson devices is still unknown at frequencies of the order of $\omega_{01}$, and their role in the relaxation of a Josephson qubit has not been clearly established. Note that a recent work on a CPB qubit[28] suggests that they might contribute significantly to relaxation.

Another cause of charge noise is the finite impedance $Z_g$ (see Fig. 5) of the Gate Line (GL), which can be treated as a set of harmonic oscillators coupled to $\widehat{N}$. As seen from the pure Josephson element of the CPB (junction capacitance not included), the gate circuit is equivalent[3] to an effective impedance $Z_{eq}$ in series with a voltage source $\kappa_g V_g$, with $\kappa_g = C_g/C_\Sigma$. In the weak coupling limit $\kappa_g \ll 1$ and for all relevant frequencies, one has $\mathrm{Re}[Z_{eq}] \simeq \kappa_g^2 \mathrm{Re}[Z_g]$. At thermal equilibrium, the contribution of the gate line to $N_g$ fluctuations is characterized by the spectral density

$$S_{Ng}^{GL}(\omega) \simeq \kappa_g^2 \frac{\hbar^2 \omega}{E_C^2} \frac{\mathrm{Re}[Z_g(\omega)]}{R_k} \left[ 1 + \coth\left( \frac{\hbar \omega}{2 k_B T} \right) \right] \,, \tag{5}$$

where $R_k = h/e^2 \simeq 26$ k$\Omega$. Using the parameters previously mentioned, we find $Sc_{Ng}^{GL}(|\omega| < 2\pi \times 10$ MHz$) \simeq (30 \times 10^{-9})^2/(\mathrm{rad/s})$ at low frequency, and $S_{Ng}^{GL}(\omega) \simeq (1 - 3 \times 10^{-9})^2/(\mathrm{rad/s})$ in the $6-17$ GHz frequency range. Finally, the out-of-equilibrium noise generated by the DC gate voltage source is fully filtered by the line and does not contribute to decoherence. The conclusion of this analysis is that the background charge noise dominates $S_{Ng}^{GL}(\omega)$ at low frequency.

### 3   Noise in $\delta$: Magnetic flux noise and readout circuitry.

The last source of decoherence encountered is the noise of the superconducting phase $\delta$. One contribution to this noise source is the magnetic flux noise threading the quantronium loop. It is however negligible because the external flux is shielded by a superconducting aluminum cylinder surrounding the sample-holder, and because the coupling to the flux coil is weak $\kappa_L = M E_J/\varphi_0^2 \ll 1$.

A second phase noise source arises from the magnetic vortices moving in the superconducting electrodes of the device. Taking the width $\ell$ of the aluminum lines used in this work, the depinning field of these vortices[27,29] $B_m \simeq \Phi_0/\ell^2$ is of order 50 mT, a value two orders of magnitude larger than the maximum field we apply, which suggests that vortices should be pinned. Nevertheless, many experiments on SQUIDs have shown that an extra flux noise whose origin is unknown, and which does not depend on the temperature below a few 100 mK[30], is always present with a spectral density

$Sc_{\delta/2\pi}^{micro}(|\omega| < 2\pi \times 1 \text{ kHz}) \sim (10 \times 10^{-6})^2/|\omega|$.

Finally, the readout circuitry also induces phase fluctuations, due to the admittance $Y_R$ (see Fig. 5) in parallel with the pure Josephson element of the readout junction, and due to the out-of-equilibrium noise of the Arbitrary Waveform Generator (AWG) used. More precisely, when a bias current $I_b < \mathcal{I}_0$ is applied to the quantronium, the effective inductance $\mathcal{L}_J \simeq (\varphi_0/\mathcal{I}_0)/\sqrt{1 - (I_b/\mathcal{I}_0)^2}$ of the readout junction converts the current noise produced by $Y_R$ into phase fluctuations characterized by the spectral density

$$S_{\delta/2\pi}^{YR}(\omega) = \frac{1}{64\pi^4}\kappa_J^2\frac{\hbar^2\omega}{E_J^2}\text{Re}[Y_\delta(\omega)] \ R_k \left[1 + \coth\left(\frac{\hbar\omega}{2k_BT}\right)\right] \ , \qquad (6)$$

where $Y_\delta(\omega) = Y_R(\omega) \ / \ |1 - \mathcal{L}_J C_J \omega^2 + j \mathcal{L}_J \omega Y_R(\omega)|^2$ and $\kappa_J = E_J/\mathcal{E}_J$. Using the parameters mentioned in the previous section, we find $Sc_{\delta/2\pi}^{YR}(|\omega| < 2\pi \times 10 \text{ MHz}) \simeq (2 \times 10^{-9})^2/(\text{rad/s})$ and $S_{\delta/2\pi}^{YR}(\omega) \simeq (20 - 80 \times 10^{-9})^2/(\text{rad/s})$ in the $6 - 17$ GHz frequency range. Then, the noise spectrum of our AWG is flat up to 200 MHz, and corresponds to a spectral density $Sc_{\delta/2\pi}^{AWG}(|\omega|) \simeq (15 \times 10^{-9}/\cos\gamma)^2/(\text{rad/s})$ that depends on the average phase $\gamma$ across the readout junction. The conclusion of this analysis is that the phase noise is dominated at low frequency by local sources close to the junction loop, and at the qubit frequency by the contribution of the biasing circuitry.

## III.  THEORETICAL DESCRIPTION OF DECOHERENCE

We now consider the dynamics of a qubit from a general point of view in two situations: the free evolution and the evolution driven by a sinusoidal excitation.

In the first case, after initial preparation in a coherent superposition of the two qubit states, the effective spin precesses freely under the influence of the static field $\vec{H}_0$, set by the control parameters $\lambda_0$, and of its classical and quantum fluctuations, set by the fluctuations $\delta\lambda$. One distinguishes two time scales, the depolarization time $T_1$ (dominated at low temperatures by the relaxation to the ground state) for the decay of the diagonal $Z$ component of the spin density matrix, and the decay time $T_2$ of the off-diagonal part, which is the qubit coherence time. As described in the experimental section IV, the time $T_2$ is inferred from the decay of Ramsey oscillations in a two $\pi/2$ pulse experiment. These Ramsey oscillations are the equivalent of the free induction decay in NMR[31]. Note that this decay can be non-exponential, the time $T_2$ being then defined by a decay by the factor $\exp(-1)$. In a modified version of the Ramsey experiment, an extra $\pi$ pulse is applied in the middle of the sequence in order to perform a Hahn echo experiment[31]. The decay time $T_{Echo}$ of this echo is longer than $T_2$, and the enhancement factor provides important information on the spectral density of the noise mechanisms.

In the second case of driven evolution, the decay of the spin density matrix is investigated in the rotating frame. Experimentally, this decay is obtained from spin-locking signals[31] and from Rabi oscillations. It is shown that time scales $\tilde{T}_1$ and $\tilde{T}_2$, similar to $T_1$ and $T_2$, describe the dynamics in the rotating frame[31].

We first start by expanding the Hamiltonian $\widehat{H}_{qb}$ of Eq. (4) to second order in the perturbation $\delta\lambda$:

$$\widehat{H}_{qb} = -\frac{1}{2}\left[\vec{H}_0(\lambda_0) + \frac{\partial \vec{H}_0}{\partial \lambda}\delta\lambda + \frac{\partial^2 \vec{H}_0}{\partial \lambda^2}\frac{\delta\lambda^2}{2} + ...\right]\vec{\widehat{\sigma}}. \qquad (7)$$

Introducing the notations $\vec{D}_\lambda \equiv (1/\hbar)\,\partial\vec{H}_0/\partial\lambda$ and $\vec{D}_{\lambda 2} \equiv (1/\hbar)\,\partial^2\vec{H}_0/\partial\lambda^2$, one finds in the eigenbasis of $\vec{H}_0(\lambda_0)$ $\vec{\widehat{\sigma}}$:

$$\widehat{H}_{qb} = -\frac{1}{2}\hbar\left(\omega_{01}\widehat{\sigma}_z + \delta\omega_z\widehat{\sigma}_z + \delta\omega_\perp\widehat{\sigma}_\perp\right) \qquad (8)$$

where $\hbar\omega_{01} \equiv |\vec{H}_0(\lambda_0)|$, $\delta\omega_z \equiv D_{\lambda,z}\delta\lambda + D_{\lambda 2,z}\,\delta\lambda^2/2 + ...$, and $\delta\omega_\perp \equiv D_{\lambda,\perp}\delta\lambda + ....$ Here $\sigma_\perp$ denotes the transverse spin components [i.e., the last term in Eq. (8) may include both $\sigma_x$ and $\sigma_y$]. We write explicitly only the terms in the expansion that dominate decoherence (as will become clear below). These coefficients $D$ are related to the derivatives of $\omega_{01}(\lambda)$:

$$\frac{\partial\omega_{01}}{\partial\lambda} = D_{\lambda,z}\ , \qquad (9)$$

and

$$D_{\lambda 2,z} = \frac{\partial^2\omega_{01}}{\partial\lambda^2} - \frac{D_{\lambda,\perp}^2}{\hbar\omega_{01}} \qquad (10)$$

As discussed below, $\partial\omega_{01}/\partial\lambda$ and $\partial^2\omega_{01}/\partial\lambda^2$ are sufficient to treat the low-frequency noise whereas the calculation of the depolarization rates involves $D_{\lambda,\perp}$.

The Bloch-Redfield theory[32,33] describes the dynamics of two-level systems (spins) in terms of two rates (times): the longitudinal relaxation (depolarization) rate $\Gamma_1 = T_1^{-1}$, and the transverse relaxation (dephasing) rate $\Gamma_2 = T_2^{-1}$. The dephasing process is a combination of effects of the depolarization ($\Gamma_1$) and of the so called 'pure' dephasing. The 'pure' dephasing is usually associated with the inhomogeneous broadening in ensembles of spins, but occurs also for a single spin due to the longitudinal low-frequency noise. It is characterized by the rate $\Gamma_\varphi$. These two processes combine to a rate

$$\Gamma_2 = \frac{1}{2}\Gamma_1 + \Gamma_\varphi \ . \tag{11}$$

The Bloch-Redfield approach applies only if the noise is short-correlated (e.g., white noise) and weak[31]. In more general situations the decay is non-exponential. In particular, when the 'pure' dephasing is dominated by a noise singular near $\omega \approx 0$, the decay law $\exp(-\Gamma_\varphi t)$ is replaced by other decay functions which we denote as $f_{z,\dots}(t)$ (additional indices ... describe the particular experiment). It can be shown[34] that the decays due to the depolarization and the 'pure' dephasing processes factorize, provided the high frequency noise responsible for the depolarization is regular. I.e., instead of the exponential decay $e^{-\Gamma_2 t}$ with $\Gamma_2$ from Eq. (11), one obtains the decay law $f_{z,\dots}(t)\exp(-\Gamma_1 t/2)$.

## A  Depolarization ($T_1$)

The depolarization rate $\Gamma_1 = T_1^{-1}$ is given by the sum,

$$\Gamma_1 = \ \Gamma_{Rel} + \Gamma_{Ex}, \tag{12}$$

of the relaxation rate $\Gamma_{Rel}$ and the excitation rate $\Gamma_{Ex}$. The Golden rule gives

$$\Gamma_{Rel} = \ \frac{\pi}{2}\, S_{\delta\omega_\perp}(\omega_{01}) \ = \ \frac{\pi}{2}\, D_{\lambda,\perp}^2 S_\lambda(\omega_{01}) \ , \tag{13}$$

$$\Gamma_{Ex} = \frac{\pi}{2} S_{\delta\omega_\perp}(-\omega_{01}) = \frac{\pi}{2} D_{\lambda,\perp}^2 S_\lambda(-\omega_{01}) . \tag{14}$$

Thus

$$\Gamma_1 = \pi S s_{\delta\omega_\perp}(\omega_{01}) = \pi D_{\lambda,\perp}^2 S s_\lambda(\omega_{01}) . \tag{15}$$

This result holds irrespective of the statistics of the fluctuations; in lowest order of the perturbation theory in $D_{\lambda,\perp}$ the rates are expressed through the correlator $S_\lambda$. This approximation is sufficient when the noise is weak enough with a smooth spectrum at the transition frequency $\omega_{01}$ on the scale of the relaxation rate $\Gamma_1$. At low temperatures $k_{\rm B}T \ll \hbar\omega_{01}$ the excitation rate $\Gamma_E$ is exponentially suppressed and $\Gamma_1 \approx \Gamma_R$.

### B  'Pure' dephasing

#### 1  Linear coupling

First, we analyze a noise source coupled linearly (and longitudinally) to the qubit, i.e. $\partial\omega_{01}/\partial\lambda = D_{\lambda,z} \neq 0$. The Bloch-Redfield theory gives for the 'pure' dephasing rate

$$\Gamma_\varphi = \pi S_{\delta\omega_z}(\omega = 0) = \pi D_{\lambda,z}^2 S_\lambda(\omega = 0) = \pi D_{\lambda,z}^2 S c_\lambda(\omega = 0). \tag{16}$$

This result is of the Golden Rule type [similar to Eq. (15)] and is meaningful if the noise power $S c_\lambda$ is regular near $\omega \approx 0$ up to frequencies of order $\Gamma_\varphi$.

A more elaborate analysis is needed when the noise spectral density is singular at low frequencies. In this subsection we consider Gaussian noise. The random phase accumulated at time $t$:

$$\Delta\phi = D_{\lambda,z} \int\limits_0^t dt' \delta\lambda(t')$$

is then Gaussian-distributed, and one can calculate the decay law of the free induction (Ramsey signal) as $f_{z,R}(t) = \langle \exp(i\Delta\phi) \rangle = \exp(-(1/2)\langle \Delta\phi^2 \rangle)$. This gives

$$f_{z,R}(t) = \exp\left[ -\frac{t^2}{2} D_{\lambda,z}^2 \int_{-\infty}^{+\infty} d\omega S_\lambda(\omega) \, \mathrm{sinc}^2 \frac{\omega t}{2} \right] , \tag{17}$$

where $\operatorname{sinc} x \equiv \sin x / x$.

In an echo experiment, the phase acquired is the difference between the two free evolution periods:

$$\Delta\phi_E = -\Delta\phi_1 + \Delta\phi_2 = -D_{\lambda,z} \int\limits_0^{t/2} dt' \delta\lambda(t') + D_{\lambda,z} \int\limits_{t/2}^{t} dt' \delta\lambda(t') \,, \qquad (18)$$

so that

$$f_{z,E}(t) = \exp\left[ -\frac{t^2}{2} D_{\lambda,z}^2 \int_{-\infty}^{+\infty} d\omega S_\lambda(\omega) \sin^2 \frac{\omega t}{4} \operatorname{sinc}^2 \frac{\omega t}{4} \right] . \qquad (19)$$

*$1/f$ spectrum:* Here and below in the analysis of noise with $1/f$ spectrum we assume that the $1/f$ law extends in a wide range of frequencies limited by an infrared cut-off $\omega_{\mathrm{ir}}$ and an ultraviolet cut-off $\omega_c$:

$$S_\lambda(\omega) = A/|\omega|, \qquad \omega_{\mathrm{ir}} < |\omega| < \omega_c \,. \qquad (20)$$

The infrared cutoff $\omega_{\mathrm{ir}}$ is usually determined by the measurement protocol, as discussed further below. The decay rates typically depend only logarithmically on $\omega_{\mathrm{ir}}$, and the details of the behavior of the noise power below $\omega_{\mathrm{ir}}$ are irrelevant to logarithmic accuracy. For most part of our analysis, the same remark applies to the ultra-violet cut-off $\omega_c$. However, for some specific questions considered below, frequency integrals may be dominated by $\omega \gtrsim \omega_c$, and thus the detailed behavior near and above $\omega_c$ ('shape' of the cut-off) is relevant. We will refer to an abrupt suppression above $\omega_c$ ($S(\omega) \propto \theta(\omega_c - |\omega|)$) as a 'sharp cut-off', and to a crossover at $\omega \sim \omega_c$ to a faster decay $1/\omega \to 1/\omega^2$ (motivated by modelling of the noise via a set of bistable fluctuators, see below), as a 'soft cut-off'.

For $1/f$ noise, at times $t \ll 1/\omega_{ir}$, the free induction (Ramsey) decay is dominated by the frequencies $\omega < 1/t$, i.e., by the quasistatic contribution[3], and ref. (17) reduces to:

$$f_{z,R}(t) = \exp\left[ -t^2 D_{\lambda,z}^2 A \left( \ln\frac{1}{\omega_{ir} t} + O(1) \right) \right] . \qquad (21)$$

The infrared cutoff $\omega_{\mathrm{ir}}$ ensures the convergence of the integral.

For the echo decay we obtain

$$f_{z,E}(t) = \exp\left[ -t^2 D_{\lambda,z}^2 A \cdot \ln 2 \right] \ . \qquad (22)$$

The echo method thus only increases the decay time by a logarithmic factor. This limited echo efficiency is due to the high frequency tail of the $1/f$ noise.

*Static case:* In many cases, the contribution of low frequencies $\omega \ll 1/t$ dominates the 'pure' dephasing. This happens when the noise spectrum is strongly peaked at low frequencies [cf. Eq. (21)], in particular when it has a sufficiently low ultraviolet cutoff frequency $\omega_c$. This simple regime pertains to the quantronium.

To fix the terminology we use here: under certain conditions we use the 'static approximation' characterized by an effective distribution $P(\delta\lambda)$, for which the noisy control parameter $\lambda$ is considered as constant during each pulse sequence. This approach allows for a direct evaluation of the Ramsey decay function $f_{z,R}(t)$. In the relevant cases of linear or quadratic coupling to the fluctuations, the decay function $f_{z,R}(t)$ is the Fourier or Fresnel-type transform of the distribution $P(\delta\lambda)$, respectively. Since the static approximation would yield no decay for the echoes, the calculation of the echo decay function $f_{z,E}(t)$ requires a 'quasistatic approximation' that takes into account variations within each pulse sequence. A noise with an ultraviolet cutoff frequency $\omega_c$ can be considered as quasistatic on time scales shorter than $\omega_c^{-1}$. The relevant results obtained in refs.[35–37] are given below.

In the static approximation, the contribution of low frequencies $\omega \ll 1/t$ to the integral in Eq. (17) is evaluated using the asymptotic value $\mathrm{sinc}(\omega t/2) \approx 1$:

$$f_{z,R}^{\mathrm{stat}}(t) = \exp\left[-\frac{t^2}{2}\, D_{\lambda,z}^2 \sigma_\lambda^2\right],\tag{23}$$

where $\sigma_\lambda^2 = \int_{-\infty}^{+\infty} d\omega\, S_\lambda(\omega)$ is the dispersion of $\delta\lambda$.

For $1/f$ noise, $S_\lambda = (A/|\omega|)\theta(\omega_c - |\omega|)$, we obtain $\sigma_\lambda^2 = 2A\ln(\omega_c/\omega_{ir})$. The result is only logarithmically sensitive to the value of the ultraviolet cutoff $\omega_c$ and to the specific functional form of the suppression of noise at high $\omega \gtrsim \omega_c$. The static approximation is sufficient for the evaluation of the dephasing rate if, e.g., the latter indeed exceeds the ultraviolet cutoff $\omega_c$, i.e., $D_{\lambda,z}^2\, A\,\ln(\omega_c/\omega_{ir}) \gg \omega_c^2$.

In the opposite limit, for the wide-band $1/f$ noise at $t > 1/\omega_c$, the contribution of frequencies $\omega \ll 1/t$ is also given by Eq. (23) [cf. Eqs.(17) and (21)]. In this case, however, $\sigma_\lambda^2$ in Eq. (23) depends logarithmically on time: $\sigma_\lambda^2 = D_{\lambda,z}^2 \, A \ln(1/\omega_{ir} t)$. This contribution dominates the decay of $f_{z,R}(t)$.

In general, for (quasi-)static noise with distribution function $P(\delta\lambda)$ the 'static' approximation yields the Ramsey decay,

$$f_{z,R}^{\text{stat}}(t) = \int d(\delta\lambda) P(\delta\lambda) \, e^{i D_{\lambda,z} \, \delta\lambda \, t} \,, \tag{24}$$

which is the Fourier transform of $P(\delta\lambda)$.

Let us now analyze the echo decay. For $1/f$ noise with a low $\omega_c$ the integral in Eq. (19) over the interval $\omega \lesssim \omega_c$ is dominated by the upper limit. This indicates that the specific behavior at $\omega \gtrsim \omega_c$ is crucial. For instance, in the case of a sharp cutoff $(S = (A/|\omega|)\theta(\omega_c - \omega))$ we obtain

$$f_{z,E}(t) = \exp\left(-\frac{1}{32} D_{\lambda,z}^2 \, A \, \omega_c^2 \, t^4\right). \tag{25}$$

However, if the $1/f$ behavior for $\omega < \omega_c$ crosses over to a faster decay $\propto 1/\omega^2$ at $\omega > \omega_c$ (as one would expect when the noise is produced by a collection of bistable fluctuators with Lorentzian spectra, cf.[38,39,41]) then the integral in Eq. (19) is dominated by frequencies $\omega_c < \omega < 1/t$, and we find: $\ln f_{z,E}(t) \propto D_{\lambda,z}^2 \, A \, \omega_c \, t^3$. In either case, one finds that the decay is slower than for $1/f$ noise with a high cutoff $\omega_c > D_{\lambda,z} A^{1/2}$: the exponent involved in the decay function is indeed reduced by a factor $\sim (\omega_c t)^2$ or $\omega_c t$, respectively.

### 2   Quadratic coupling

At the optimal working point, the first-order longitudinal coupling $D_{\lambda,z}$ vanishes. Thus, to first order, the decay of the coherent oscillations are determined by the relaxation processes and one expects $\Gamma_2 = \Gamma_1/2$ from Eq. (11). However, it turns out that due to a singularity at low frequencies the second-order contribution of the longitudinal noise can be comparable, or even dominate over $\Gamma_1/2$. To evaluate this contribution, one has to calculate

$$f_z(t) = \left\langle \exp\left(i \, \frac{1}{2} \, \frac{\partial^2 \omega_{01}}{\partial \lambda^2} \int_0^t \chi(\tau) \, \delta\lambda^2(\tau) d\tau\right)\right\rangle. \tag{26}$$

Equation (26) can be used for the analysis of the free induction decay (Ramsey signal) if one sets $\chi(\tau) = 1$, and for the investigation of the echo-signal decay using $\chi(\tau < t/2) = -1$ and $\chi(\tau > t/2) = 1$.

$1/f$ *noise:* The free induction decay for the $1/f$ noise with a high cutoff $\omega_c$ (the highest energy scale in the problem) has been analyzed in Ref.[35]. The decay law can be approximated[40] by the product of the low-frequency ($\omega < 1/t$, quasi-static) and the high-frequency ($\omega > 1/t$) contributions: $f_{z,R}(t) = f_{z,R}^{\text{lf}}(t) \cdot f_{z,R}^{\text{hf}}(t)$. The contribution of low frequencies is given by (cf.[35–37]):

$$f_{z,R}^{\text{lf}}(t) = \frac{1}{\sqrt{1 - i \frac{\partial^2 \omega_{01}}{\partial \lambda^2} \sigma_\lambda^2 t}} \ . \tag{27}$$

For $1/f$ noise with variance of the low-frequency fluctuations $\sigma_\lambda^2 = 2A \ln(1/\omega_{\text{ir}} t)$, this contribution is

$$f_{z,R}^{\text{lf}}(t) = \frac{1}{\sqrt{1 - 2i \frac{\partial^2 \omega_{01}}{\partial \lambda^2} t A \ln \frac{1}{\omega_{ir} t}}} \ . \tag{28}$$

It dominates at short times $t < \left[ (\partial^2 \omega_{01}/\partial \lambda^2) A/2 \right]^{-1}$. At longer times, the high-frequency contribution

$$\ln f_{z,R}^{\text{hf}}(t) \approx -t \int\limits_{\sim 1/t}^{\infty} \frac{d\omega}{2\pi} \ln \left( 1 - 2\pi i \frac{\partial^2 \omega_{01}}{\partial \lambda^2} S_\lambda(\omega) \right) \ , \tag{29}$$

takes over: when $t \gg \left[ (\partial^2 \omega_{01}/\partial \lambda^2) A/2 \right]^{-1}$ we obtain asymptotically $\ln f_{z,R}^{\text{hf}}(t) \approx -(\pi/2)(\partial^2 \omega_{01}/\partial \lambda^2) At$ (provided $\omega_c \gg \pi (\partial^2 \omega_{01}/\partial \lambda^2) A$). Otherwise the quasistatic result (27) is valid at all relevant times. One can also evaluate the pre-exponential factor in the long-time decay. This pre-exponent decays very slowly (algebraically) but differs from 1 and thus shifts the level of $f_{z,R}(t)$[42].

Note that the experimentally monitored quantity is a spin component, say $\langle \sigma_x \rangle$, in the rotating frame which evolves according to $\langle \sigma_x \rangle = \text{Re}[f_{z,R}(t) \, e^{i\Delta\omega t}]$, where $\Delta\omega$ is the detuning frequency. In a typical situation of interest $f_{z,R}(t)$ changes slower than the period of oscillations, and thus the envelope of the decaying oscillations is given by $|f_{z,R}(t)|$, the phase of $f_{z,R}(t)$ shifting the phase of the oscillations. In the opposite limit $\Delta\omega = 0$, the measured decay curve reproduces the real part of $f_{z,R}(t)$ (the imaginary part corresponds to $\sigma_y$ and can also be measured).

*Static case:* In the quasi-static case, i.e., when the cutoff $\omega_c$ is lower than $1/t$ for all relevant times, the Ramsey decay is simply given by the static contribution (27). At all relevant times the decay is algebraic and the crossover to the exponential law is not observed. More generally, in the static approximation with a distribution $P(\delta\lambda)$, the dephasing law is given by the Fresnel-type integral transform:

$$f_{z,R}^{\text{st}}(t) = \int d\left(\delta\lambda\right) P(\delta\lambda)\, e^{i\frac{1}{2}\frac{\partial^2\omega_{01}}{\partial\lambda^2}\delta\lambda^2\, t}\,, \tag{30}$$

which reduces to Eq. (27) for a Gaussian $P(\delta\lambda) \propto \exp\left(-\delta\lambda^2/2\sigma_\lambda^2\right)$. In general, any distribution $P(\delta\lambda)$, finite at $\delta\lambda = 0$, yields a $t^{-1/2}$ decay for $f_{z,R}^{\text{st}}$ at long times.

For the echo decay and Gaussian quasi-static noise in $\lambda$ we obtain

$$f_{z,E}(t) = \frac{1}{\sqrt{1 + \left(\frac{\partial^2\omega_{01}}{\partial\lambda^2}\right)^2 \sigma_\lambda^2 \int_{-\infty}^{+\infty} d\omega \left(\frac{\omega t}{4}\right)^2 S_\lambda(\omega)\, t^2}}\,, \tag{31}$$

where we assumed that the frequency integral converges at $|\omega| \ll 1/t$. This is the case, e.g., if $S_\lambda(\omega)$ has a sharp cut-off at $\omega_c \ll 1/t$. For $1/f$ noise, $S_\lambda = (A/|\omega|)\theta(\omega_c - |\omega|)$ with $\omega_c \ll 1/t$, Eq. (31) yields

$$f_{z,E}(t) = \frac{1}{\sqrt{1 + \frac{1}{16}\left(\frac{\partial^2\omega_{01}}{\partial\lambda^2}\right)^2 \sigma_\lambda^2 A\, \omega_c^2\, t^4}}\,. \tag{32}$$

Note that this result is sensitive to the precise form of the cut-off.

## C $1/f$ noise, one fluctuator versus many

The background charge fluctuations are induced by random redistributions of charge near, e.g., trapping and release of electrons or by random rearrangements of charged impurities. Many groups have observed this noise with a smooth $1/f$ spectrum in the frequency range from 1 Hz to 1 MHz. Occasionally, single fluctuators have been observed, with a significant fraction of the total charge noise[10]. If individual fluctuators play an important part the noise statistics is non-Gaussian[39,43]. We summarize here some of the obtained results relevant to our work.

The noise $\delta\lambda(t)$ contains contributions from all TLFs:

$$\delta\lambda(t) = \sum_n v_n \sigma_{n,z}(t). \qquad (33)$$

Every fluctuator switches randomly between two positions, denoted by $\sigma_{n,z} = \pm 1$ with rate $\gamma_n$ (for simplicity, we assume equal rates in both directions for relevant TLFs) and thus contributes to the noise power $S_\lambda = \sum_n S_n$:

$$S_n = \frac{1}{\pi} \frac{\gamma_n v_n^2}{\omega^2 + \gamma_n^2}. \qquad (34)$$

For a (longitudinally coupled to the qubit) single fluctuator the free induction (Ramsey) and the echo decays are given by

$$f_{z,R,n}(t) = e^{-\gamma_n t} \left( \cos\mu_n t + \frac{\gamma_n}{\mu_n} \sin\mu_n t \right) , \qquad (35)$$

and

$$f_{z,E,n}(t) = e^{-\gamma_n t} \left( 1 + \frac{\gamma_n}{\mu_n} \sin\mu_n t + \frac{\gamma_n^2}{\mu_n^2}(1 - \cos\mu_n t) \right) , \qquad (36)$$

where $\mu_n \equiv \sqrt{(D_{\lambda,z} v_n)^2 - \gamma_n^2}$. Finally, the decay produced by all the fluctuators is just the product of the individual contributions, i.e., $f_{z,R}(t) = \Pi_n f_{z,R,n}(t)$ and $f_{z,E}(t) = \Pi_n f_{z,E,n}(t)$. If the noise is produced or dominated by a few fluctuators, the distribution of $\delta\lambda(t)$ may be strongly non-Gaussian, and the simple relation between decoherence and noise power does not hold. In this case the conditions of the central limit theorem are not satisfied. In Ref.[43], a continuous distribution of $v_n$'s and $\gamma_n$'s was considered, with a long tail of the distribution of the coupling strengths $v_n$ such that rare configurations with very large $v_n$ dominate the ensemble properties. The distribution $P(v, \gamma)$ considered in Ref.[43] is defined in the domain $[v_{min}, \infty] \times [\gamma_{min}, \gamma_{max}]$:

$$P(v, \gamma) = \frac{\xi}{\gamma v^2}. \qquad (37)$$

Let us introduce the parameter $v_{max}^{typ} \equiv N v_{min}$, which gives the typical value of the strongest (closest) fluctuator. Normalization to $N$ fluctuators requires that $\xi = v_{max}^{typ} / \ln(\gamma_{max}/\gamma_{min})$.

For this distribution any quantity whose average value (that is integrals over $v$'s and $\gamma$'s) is dominated by TLFs with $v \gtrsim v_{\max}^{\mathrm{typ}}$ [43] is not self averaging, i.e. has considerable sample-to-sample fluctuations. The ensemble-averaged free induction decay described by[43]

$$\ln |f_{z,R}(t)| \propto -D_{\lambda,z}\,\xi\,t\,\ln(\gamma_{max}/\gamma_{min}) = -D_{\lambda,z}\,v_{\max}^{\mathrm{typ}}t \qquad (38)$$

is dominated by the fluctuators with strength of order $v \sim v_{\max}^{\mathrm{typ}}$ and is thus not self-averaging. Consequently, it does not apply quantitatively to a specific sample.

Similarly, the ensemble-averaged echo signal is given by[43]

$$\ln |f_{z,E}(t)| \propto -D_{\lambda,z}\,\xi\,\gamma_{max}t^2 \quad \text{for} \quad t < \gamma_{max}^{-1}$$
$$\ln |f_{z,E}(t)| \propto -D_{\lambda,z}\,\xi\,t\,[\ln(\gamma_{max}t) + O(1)] \quad \text{for} \quad t > \gamma_{max}^{-1}\,. \qquad (39)$$

The situation depends on whether $D_{\lambda,z}\xi > \gamma_{max}$ or $D_{\lambda,z}\xi < \gamma_{max}$. In the former the dephasing is 'static' (i.e., it happens on a time scale shorter than the flip time of the fastest fluctuators, $1/\gamma_{max}$) and first line of Eq. (39) applies. The decay is self-averaging because it is dominated by many fluctuators with strength $v \approx \sqrt{\xi\,\gamma_{max}/D_{\lambda,z}} < \xi < v_{\max}^{\mathrm{typ}}$. In the opposite regime $D_{\lambda,z}\xi < \gamma_{max}$ the dephasing is due to multiple flips of the fluctuators and the the second line of Eq. (39) applies. In this case, the decay is dominated by a small number of fluctuators with strength $v \approx \xi$, which is smaller than $v_{\max}^{\mathrm{typ}}$ only by a logarithmic factor, and sample-to-sample fluctuations are strong.

### D  Decoherence during driven evolution

In the presence of a harmonic drive $2\omega_{R0}\cos(\omega t)\widehat{\sigma}_x$, the Hamiltonian reads

$$\widehat{H} = -\frac{1}{2}\hbar\,[\omega_{01}\widehat{\sigma}_z + \delta\omega_z\widehat{\sigma}_z + \delta\omega_\perp\widehat{\sigma}_\perp + 2\omega_{R0}\cos(\omega t)\widehat{\sigma}_x]\,. \qquad (40)$$

The qubit dynamics is conveniently described in the frame rotating with the driving frequency $\omega$, and a new eigenbasis $\left\{\left|\widetilde{0}\right\rangle, \left|\widetilde{1}\right\rangle\right\}$ is defined by the total static fictitious field composed of the vertical component given by the detuning $\Delta\omega = \omega_{01} - \omega$ and the horizontal ($x$) component $\omega_{R0}$. I.e., the static part of the Hamiltonian in the rotating frame reads

$$H_{\mathrm{st}} = -\frac{1}{2}\hbar\left[\Delta\omega\,\widehat{\sigma}_z + \omega_{R0}\,\widehat{\sigma}_x\right]. \tag{41}$$

The length of the total field is $\omega_R = \sqrt{\omega_{R0}^2 + \Delta\omega^2}$ and it makes an angle $\eta$ with the $z$-axis: $\Delta\omega = \omega_R\cos\eta$, $\omega_{R0} = \omega_R\sin\eta$. The evolution of the spin is a rotation around the field at the Rabi precession frequency $\omega_R$. Like in the case of free evolution, decoherence during driven evolution involves the phenomena of relaxation and dephasing: one defines a relaxation time $\tilde{T}_1$ and a coherence time $\tilde{T}_2$ analogous to $T_1$ and $T_2$, which correpond to the decay of the longitudinal and of the transverse part of the density matrix[31] in the new eigenbasis, respectively. First, as a reference point, we present the Golden-Rule-type results which are valid if all the noises are smooth at frequencies near $\omega = 0$, $\omega_R$, and $\omega_{01}$. Analyzing which parts of the fluctuating fields $\delta\omega_z$ and $\delta\omega_\perp$ are longitudinal and transverse with respect to the total field $\omega_R$ in the rotating frame, and taking into account the frequency shifts due to the transformation to the rotating frame we obtain

$$\tilde{\Gamma}_1 = \sin^2\eta\,\Gamma_\nu + \frac{1 + \cos^2\eta}{2}\,\Gamma_1\ , \tag{42}$$

where $\Gamma_\nu \equiv \pi S_{\delta\omega_z}(\omega_R)$ is the spectral density at the Rabi frequency. We have disregarded the difference in the noise power $S_{\delta\omega_\perp}$ at frequencies $\omega_{01}$ and $\omega_{01} \pm \omega_R$, which allows us to use the depolarization rate $\Gamma_1$ from Eq. (15). We do, however, distinguish between $\Gamma_\nu$ and $\Gamma_\varphi = \pi S_{\delta\omega_z}(\omega = 0)$ in order to, later, analyze a noise spectrum singular at $\omega \approx 0$.

For the dephasing rate we again have the relation

$$\widetilde{\Gamma}_2 = \frac{1}{2}\widetilde{\Gamma}_1 + \widetilde{\Gamma}_\varphi, \tag{43}$$

where

$$\tilde{\Gamma}_\varphi = \Gamma_\varphi\cos^2\eta + \frac{1}{2}\Gamma_1\sin^2\eta\ . \tag{44}$$

As a result we obtain

$$\tilde{\Gamma}_2 = \frac{3 - \cos^2\eta}{4}\Gamma_1 + \Gamma_\phi\cos^2\eta + \frac{1}{2}\,\Gamma_\nu\sin^2\eta\ . \tag{45}$$

The derivation of these expressions is simplified if one notes that due to the fast rotation the high frequency transverse noise $S_{\delta\omega_\perp}(\omega \approx \omega_{01})$ is effectively mixed to low frequencies $\lesssim \omega_R$. In the rotating frame it effectively reduces to 'independent' white noises both in the $x$ and $y$ directions with amplitudes $\delta\omega_\perp/\sqrt{2}$ and corresponding noise powers $S_{\delta\omega_\perp}(\omega \approx \omega_{01})/2$. Only the noise along the $x$ axis (its longitudinal component with factor $\sin^2\eta$) contributes to $\tilde{\Gamma}_\varphi$ (the noise along the $y$ axis is purely transverse).

Note the limiting behavior of the rates: at zero detuning, one has $\cos\eta = 0$ and $\tilde{\Gamma}_2 = \frac{3}{4}\Gamma_1 + \frac{1}{2}\Gamma_\nu$, whereas at large detuning compared to the Rabi frequency, $\cos\eta = 1$, and $\tilde{\Gamma}_2 = \frac{1}{2}\Gamma_1 + \Gamma_\phi$: one recovers thus the decoherence rate $\Gamma_2$ of the free evolution.

For a noise spectrum which is singular at $\omega = 0$ ($1/f$ noise) we no longer find the exponential decay. The simplest case is when the Rabi frequency is high enough so that one still can use the rate $\Gamma_\nu$ and the associated exponential decay. We consider here only this regime. Then one should combine the exponential decay associated with the rates $\Gamma_1$ and $\Gamma_\nu$ with the non-exponential one substituting the rate $\Gamma_\varphi$. For the decay of the Rabi oscillations we obtain

$$f_{Rabi}(t) = f_{z,\cos\eta}(t) \cdot \exp\left(-\frac{3-\cos^2\eta}{4}\Gamma_1\,t - \frac{1}{2}\,\Gamma_\nu \sin^2\eta\,t\right)\;, \qquad (46)$$

where $f_{z,\cos\eta}(t)$ is given by one of the decay laws derived in the preceding sections (depending on whether the coupling is linear or quadratic, and whether the statistics is Gaussian or not) with the noise $\delta\omega_z$ substituted by $\cos\eta\,\delta\omega_z$. That is, in the linear case, we have to substitute $D_{\lambda,z} \to \cos\eta\,D_{\lambda,z}$, while in the quadratic case $(\partial^2\omega_{01}/\partial\lambda^2) \to \cos\eta\,(\partial^2\omega_{01}/\partial\lambda^2)$.

### E  Application to the Quantronium sample used in this work

As already mentioned in section II B, the parameters of the qubit $E_J = 0.87\;k_B$ K, $E_C = 0.66\;k_B$ K were measured by fitting the spectroscopic data $\omega_{01}(N_g, \delta)$ (see Fig. 11) with a numerical diagonalization of the Hamiltonian $\widehat{H}_{CPB}$. This fit gives an upper limit for the asymmetry of the qubit junctions, $d < 13$ %. By using another property[16] this value was estimated as $d \sim 4\%$. From $E_J$, $E_C$, and $d$, the numerical values of the $D_\lambda$'s introduced above were calculable exactly, as a function of the working point $(\delta, N_g)$. Nevertheless, since we have characterized decoherence only along the two segments $\delta/(2\pi) \in [-0.3, +0.3]$, $N_g = 1/2$ and $\delta = 0$, $N_g - 1/2 \in [-0.1, +0.1]$ in the $(\delta, N_g)$ plane, we only give below simple expressions that approximate $\omega_{01}$, $D_\lambda$ and $\partial^2 \omega_{01}/\partial \lambda^2$ with a $\pm 3\%$ accuracy in the range of parameters explored experimentally. We have thus for the transition frequency

$$\omega_{01}(\delta, N_g = 1/2) \simeq \left[103 - 425 \left(\delta/2\pi\right)^2\right] \times 10^9 \text{rad/s}, \tag{47}$$

$$\omega_{01}(\delta = 0, N_g) \simeq \left[103 + 145 \left(N_g - 1/2\right)^2\right] \times 10^9 \text{ rad/s}, \tag{48}$$

which lead for the longitudinal coefficients to:

$$D_{\delta/2\pi, z}(\delta = 0 \text{ or } N_g = 1/2) = \frac{1}{2e}(i_1 - i_0) \simeq -850 \frac{\delta}{2\pi} \times 10^9 \text{ rad/s}, \tag{49}$$

$$\frac{\partial^2 \omega_{01}}{\partial(\delta/2\pi)^2} \simeq -850 \times 10^9 \text{ rad/s}, \tag{50}$$

$$D_{N_g, z}(\delta = 0 \text{ or } N_g = 1/2) = -\frac{2E_C}{\hbar}(\langle 1| \widehat{\mathbf{N}} |1\rangle - \langle 0| \widehat{\mathbf{N}} |0\rangle) \simeq +290 \left(N_g - \frac{1}{2}\right) \times 10^9 \text{ rad/s}, \tag{51}$$

$$\frac{\partial^2 \omega_{01}}{\partial N_g^2} \simeq +290 \times 10^9 \text{ rad/s}, \tag{52}$$

where $i_0$ and $i_1$ are the average currents $\langle \widehat{i} \rangle$ in the two states. Note that $D_{N_g, z}$ vanishes at $N_g = 1/2$ for all $\delta$ so that a gate microwave pulse corresponds to a purely transverse perturbation of the Hamiltonian. Consequently, the perturbed Hamiltonian of Eq. 40 does apply exactly to the quantronium at $N_g = 1/2$, where the measurements were performed. At other values of $N_g$, Eq. 40 would nevertheless be a good approximation. For critical current noise, the coupling coefficient

$$D_{\delta E_J/E_J, z}(\delta = 0) \simeq +85 \times 10^9 \text{ rad/s} \tag{53}$$

$$D_{\delta E_J/E_J, z}(N_g = 1/2) \simeq \left[+85 - 240 \left(\frac{\delta}{2\pi}\right)^2\right] \times 10^9 \text{ rad/s} \tag{54}$$

is maximal at the optimal working point $P_0$. Expressed in the same way, the transverse coefficients $D_{\lambda,\perp}$ are

$$D_{\delta/2\pi,\perp}(\delta = 0 \text{ or } N_g = 1/2) = \frac{1}{e}\left|\langle 0|\,\hat{\mathbf{i}}\,|1\rangle\right| \simeq 380 \; d \; [1 + 6.0 \; (\frac{\delta}{2\pi})^2] \; \times 10^9 \text{ rad/s}, \quad (55)$$

$$D_{Ng,\perp}(\delta = 0 \text{ or } N_g = 1/2) = \frac{4E_C}{\hbar}\left|\langle 0|\,\hat{\mathbf{N}}\,|1\rangle\right| = 193 \; \times 10^9 \text{ rad/s}, \quad (56)$$

and

$$D_{\delta E_J/E_J,\perp}(\delta = 0 \text{ or } N_g = 1/2) = \frac{E_J\cos(\delta/2)}{\hbar}\left|\langle 0|\cos\widehat{\theta}\;|1\rangle\right| = 54 \left|N_g - \frac{1}{2}\right| \; \times 10^9 \text{ rad/s}. \quad (57)$$

Finally, note that the cross derivative $\partial^2\omega_{01}/\partial\delta\partial N_g$ was found to be equal to zero along the two segments mentioned above.

## IV. EXPERIMENTAL CHARACTERIZATION OF DECOHERENCE DURING FREE EVOLUTION

In order to characterize decoherence in our quantronium sample and to compare with the theoretical predictions, we have measured the characteristic decay times of the diagonal ($T_1$) and non diagonal ($T_2$, $T_E$) parts of the density matrix of the qubit during its free evolution. These measurements were repeated at different working points $P$ located along the lines $\delta = 0$ and $N_g = 1/2$, as mentioned above. We describe the different experimental protocols that were used, the results, and their interpretation.

### A  Longitudinal relaxation time, $T_1$.

Relaxation of the longitudinal polarization is inferred from the decay of the switching probability $p$ after a $\pi$ pulse has brought the qubit to state $|1\rangle$. More precisely, a sequence that consists of a $\pi$ pulse, a variable delay $t$, and a readout pulse is repeated to determine $p(t)$. An example of the relaxation curve, measured at the working point $P_0$, is shown in the inset of Fig. 6. As predicted, the relaxation is exponential, with an absolute discrepancy between $p(t)$ and the fit being always smaller than 2 %. The relaxation time $T_1$, varies with the working point as shown in Fig. 6: $T_1$ is about 0.5 $\mu s$ in the vicinity of $P_0$ (which is 3 times shorter than in a previous experiment[2]) and shows rapid variations away from $P_0$ in the phase direction. Now, it is interesting to note that in the parameter range explored, the matrix element $D_{Ng,\perp}$ of Eq. (56) is approximately constant and that the matrix element $D_{\delta/2\pi,\perp}$ of Eq. (55) varies smoothly by a factor of only 2 with $\delta$. Consequently, the measured variation of $T_1$ reflects quite directly the variation with frequency of the density of environmental modes available for absorbing one photon $\hbar\omega_{01}$ from the qubit through the $\delta$ and $N_g$ channels. Noting from Eq. (57) that the noise on $E_J$ cannot induce relaxation of the qubit along the line $N_g = 1/2$, a natural question arises: can the measured relaxation rates be fully accounted for by the circuit alone, i.e. by $Z_g$ and $Y_R$ (see Fig. 5)? We have calculated from Eq. (15) and from the noise spectra (5,6) of $Z_g$ and $Y_R$, values of $T_1$ at $P_0$ of about 2 $\mu s$ and $3-6$ $\mu s$, respectively. The combined effect of the two sub-circuits gives thus $T_1 \sim 1-1.5$ $\mu s$, which is approximately 2-3 times longer that the measured value. We conclude that if our estimates of the circuit impedances above 14 GHz and of the asymmetry coefficient $d \sim 3-4$ % are correct, a large part of the relaxation has to be attributed to microscopic environmental modes, non uniformly distributed in frequency. Note however that estimating the impedances as seen from the qubit above 14 GHz with an accuracy better than a factor 2, is difficult.

### B  Transversal relaxation time or coherence time, $T_2$.

#### 1  $T_2$ measurement from Ramsey fringes

Characterizing decoherence during the free evolution of a qubit can be done directly by measuring the temporal decay of the average transverse polarization of its effective spin.

With a projective readout, this information can only be obtained by repeating a sequence which consists in preparing first a particular state with a non zero transverse polarization, letting the spin evolve freely during a time $\Delta t$, and then reading one of its transverse components. Starting from state $|0\rangle$, the simplest experiment would consist in applying a $\pi/2$ pulse to align the spin along the $X$ axis of the Bloch sphere, and for measurement projecting it onto $X$ after the desired free evolution. Such an experiment is not possible with the quantronium, which is projected onto the $Z$ axis at readout. The phase $\varphi$ accumulated during the free precession has thus to be converted into a polarization along $Z$, which can be done by applying a second $\pi/2$ pulse. The two $\pi/2$ pulses form the so-called Ramsey sequence[2], which gives an oscillation of the $Z$ polarization with $\Delta t$ at the detuning frequency $\Delta\omega/2\pi$. Although choosing $\Delta\omega = 0$ gives a simple non oscillatory signal that decays in principle as $(1 + e^{-\Gamma_1 \Delta t/2}\mathrm{Re}\,[f_{z,R}(\Delta t)])/2$ (see section III), this choice is inconvenient since any residual detuning would induce a very slow oscillation that could be misinterpreted as an intrinsic decay. For that reason, we use here a $\Delta\omega$ of several tens of MHz, which is chosen because it is much larger than the decoherence rate. The rotation axis of the spin during the $\pi/2$ pulses makes an angle $\alpha = \arctan(\Delta\omega/\omega_{R0})$ with the equatorial plane of the Bloch sphere. The rotation angle of the so-called $\pi/2$ pulses is more exactly $\pi/2(1 + \epsilon)$, where $\epsilon$ is a small positive or negative correction due to two effects: First, the pulse duration is optimized at zero detuning, by maximizing the switching probability of the readout junction immediately after two adjacent $\pi/2$ pulses. This duration is then kept constant for a Ramsey experiment at finite detuning, so that ideally, $0 \leq \epsilon = \sqrt{1 + \tan(\alpha)^2} - 1 \lesssim 10^{-2}$. Second, the optimization procedure is done with a finite accuracy and $\epsilon$ can be different from this ideal value. The Ramsey oscillation $p_R$ is given by

$$p_R = \frac{1-a}{2}\left[1 + a\,e^{-\frac{\Delta t}{T_1}} + (1+a)\,e^{-\frac{\Delta t}{2T_1}}\,|f_{z,R}(\Delta t)|\cos\left(\Delta\omega\Delta t + \zeta\right)\right]\ , \qquad (58)$$

where $a = \sin^2\alpha - \sin\epsilon(1 - \sin^2\alpha)$ and $\zeta = \arctan\left[\sin\alpha\,(1 + \sin\varepsilon)\,/\cos\epsilon\right]$ are geometrical corrections. Note that, at large $\Delta t$, the envelope of the oscillations has an amplitude and a saturation value that depends on $\Delta\omega$.

Figure 7 shows two typical Ramsey signals measured at the optimal working point $P_0$ with $\omega_{R0} = 106$ MHz and $\Delta\omega = 50$ MHz. These two signals differ significantly although they were recorded the same day with the same experimental protocol: $N_g$ is first tuned so that the central frequency of the spectroscopic line is minimum and the Ramsey fringes are then recorded at a speed of 1 point per second, the longest record (middle frame of Fig. 7) taking thus 17 minutes. The relative non reproducibility between the two records is typical of what we have observed during several months of experimentation. It is attributed to the frequency drift induced by the $1/f$ charge noise. This drift is partly continuous and partly due to sudden jumps attributed to few strongly coupled charged TLFs, as mentioned in sections II and III. These sudden jumps are reversible and induce correlated phase and amplitude jumps of the Ramsey fringes, as shown by the arrows in the bottom panels of Fig. 7. The figure also shows a fit of the external envelope of the fringes to Eqs. (58) and (27), valid for a quadratic coupling to a static charge noise (this choice will be explained in section IV D). The values of $T_1$ and of the sensitivity to noise Eq. (52) being known, the fitting parameters are the amplitude and saturation value of the fringes, and the variance $\sigma_\lambda^2$ of the noise. The corresponding effective $T_2$ time is $300 \pm 50$ ns for this record, but it is found to vary in the range $200 - 300$ ns (see for instance top panel of Fig. 10) depending on our ability to set the working point precisely at $P_0$ and on the probability that the system stays at that point during a full record.

A series of Ramsey oscillations measured at different working points $P$ is shown in Fig. 8. Since $\omega_{01}$ and therefore $\omega_{R0}$ (at constant microwave amplitude) vary with $P$, the microwave frequency was varied in order to keep $\Delta\omega$ between 40 MHz and 100 MHz and the pulse duration was varied to maintain the rotation angle close to $\pi/2$. Note that the mean level and the amplitude of the oscillations vary due to these $\Delta\omega$ changes. A direct comparison between the Ramsey patterns shows that $T_2$ decreases dramatically when $P$ is moved away from $P_0$. More precisely, each curve gives a value $T_2(P)$ with an uncertainty of about 30%, which is plotted on Fig. 15.

### 2   $T_2$ measurement  with the 'detuning pulse' method

Probing decoherence at different working points $P$ with the Ramsey method presented above requires recalibrating for each $P$ the frequency and duration of the two $\pi/2$ pulses. Now, the $\pi/2$ pulses and the free evolution period probing decoherence, do not have to be performed at the same working point. It is thus experimentally more efficient to perform the $\pi/2$ rotations always at the optimal point $P_0$ with fixed optimized microwave pulses, and to move adiabatically to any point $P$ where decoherence is to be measured, between these pulses. This scheme, which leads also to the coherence time $T_2(P)$, is referred in the following as 'the detuning pulse method'. It has been demonstrated by moving back and forth the working point from $P_0$ to $P$ with a trapezoidal $N_g$ or $\delta$ pulse of duration $\Delta t_2$ inserted in the middle of a Ramsey sequence. Since the qubit frequency is not the same at $P$, the switching probability oscillates with $\Delta t_2$ at a new detuning frequency $\Delta \omega_2(P)$ different from $\Delta \omega$. These oscillations decay with the characteristic time $T_2(P)$. The adiabaticity criterion mentioned in section II A 1 is easily fulfilled even with a rate of change $\partial \lambda / \partial t$ as fast as $0.1/\text{ns}$. In our experiment the shortest rise/fall times $t_r$ were 10 ns and 60 ns for $N_g$ and $\delta$, respectively. This method, which is of course limited to working points $P$ where $T_2(P) \gtrsim t_r$, has been used in the ranges $|\delta| < 0.1$ and $|N_g - 1/2| < 0.05$. Examples of experimental curves are shown on Figs. 9 and 10. Each curve leads to a $T_2(P)$ value with a 50% total uncertainty, these are also shown on Fig. 15.

### 3   $T_2$ measurement from resonance line shape

When the decoherence rate becomes comparable to the Rabi frequency, time domain experiments using resonant pulses can no longer be performed and one has to operate in the frequency domain. In the linear response regime, i.e. at low microwave power, the shape of the resonance line recorded during continuous microwave excitation is simply the Fourier transform of the envelope of the free evolution decay (i.e. the Ramsey signal). One has $T_2 = k/(\pi W)$ with $W$ the resonance full width at half maximum (FWHM) and $k$ a numerical coefficient that depends on the line shape: $k = 1$ for a Lorentzian, $k = 1.6$ for a Gaussian, etc. In order to reach the linear regime, the lineshape is recorded at different decreasing microwave powers until its width saturates at the lower value. At that stage, the signal to noise ratio is usually small and the lineshape has to be averaged over a few minutes. A series of resonance lines is shown on Fig. 11, together with their positions as a function of the working point (which leads to $E_J$ and $E_C$ as previously mentioned). The rapid broadening of the line when departing from $P_0$ is clearly visible. Lineshapes at $N_g \neq 1/2$ are structured with several sub-peaks that are stable only on timescales of a few minutes. We take this again to be due to the presence of large individual charged TLF's. At $\delta \neq 0$, the lines are smoother but the low signal to noise in the linear regime does not really allow a discrimination between a Lorentzian or a Gaussian shape. We thus calculated a $T_2(P)$ using an intermediate value $k = 1.3$ and with an extra 30% uncertainty. These $T_2$'s with typical uncertainty 50% are also added to Fig. 15. Finally, the lineshape at $P_0$ is averaged over 10 minutes and is shown on Fig. 11. Its exact shape is discussed in subsection IV D.

## C    Coherence time of spin echoes -$T_E$-

In NMR[31], the spin-echo technique is a standard way to cancel the lineshape broadening of an ensemble of spins due to the spatial inhomogeneity of the magnetic field. In our case, there is a single spin (i.e. the quantronium) measured repetitively and the echo technique can compensate for a drift of the transition frequency during the time needed (about 1 s) for the repeated measurement to obtain a probability $p$. The method thus cancels a low frequency *temporal* inhomogeneity and leads to a more intrinsic coherence time $T_E > T_2$ independent of the measurement time of $p$. In practice, the spin echo sequence is a modified Ramsey sequence with an extra $\pi$ pulse placed symmetrically between the two $\pi/2$ pulses. This $\pi$ rotation around the same axis as that of the $\pi/2$ pulses makes the spin trajectory along the equator longer or shorter depending on whether $\nu_{01}$ increases or decreases. Consequently, the random phases accumulated before and after the $\pi$ pulse compensate exactly if the frequency does not change on the time scale of a sequence.

In Fig. 12, we show a series of echo signals recorded at $P_0$ by sweeping the delay $\Delta t$ between the two $\pi/2$ pulses while keeping constant the delay $\Delta t_3$ between the $\pi$ and second $\pi/2$ pulses. This protocol results in an oscillation $p(\Delta t)$ whose amplitude first decays as the usual Ramsey signal, and has then a second maximum at $\Delta t = 2\Delta t_3$. Note that at this precise echo time, the value of $p$ is an oscillation minimum. By taking advantage of the time stability of our pulse sequencer, it was possible to map directly this minimum $p_E$ by sweeping $\Delta t$ while keeping the $\pi$ pulse precisely in the middle of the sequence, as shown in Fig. 13. Ideally, at zero detuning, this mapping of $p_E$ is expected to increase as $[1 - e^{-\Gamma_1 \Delta t/2} f_{z,E}(\Delta t)]/2$ (see section III). In practise, one has once again to take into account geometric corrections due to the finite detuning, to the finite duration of the $\pi/2$ and $\pi$ pulses, and to the inaccuracy of their rotation angles. Using a generalized Bloch-Redfield approach, we find

$$
\begin{aligned}
2p_E = \Big\{ 1 - \big( a_1 + a_2 e^{-\frac{\Gamma_1 \Delta t}{2}} + a_3 e^{-\Gamma_1 \Delta t} \big) \Big\} - \\
e^{-\frac{\Gamma_1 \Delta t}{2}} \Big\{ (1 - a_4)\, f_{z,E}(\Delta t) + a_5 \mathrm{Re}\left[ e^{-\frac{i(\Delta \omega \Delta t + \xi_1)}{2}} f_{z,R}(\Delta t) \right] + \\
\Big( a_6 e^{-\frac{\Gamma_1 \Delta t}{4}} + a_7 e^{+\frac{\Gamma_1 \Delta t}{4}} \Big) \mathrm{Re}\left[ e^{-\frac{\Delta \omega \Delta t + \xi_2}{2}} f_{z,R}(\frac{\Delta t}{2}) \right] \Big\} \ , \quad (59)
\end{aligned}
$$

where the $a_i$'s are small geometrical coefficients that depend only on the angle $\alpha$ and on the errors in the microwave pulse durations. The latter terms of Eq. (59) show that on top of the expected increase of $p_E$ mentioned above, pulse imperfections induce small oscillations of $p_E$ whose damping is given by the Ramsey function $f_{z,R}$ rather than by the echo function $f_{z,E}$.

Experimental $p_E(\Delta t)$ curves recorded at $P_0$ and at different working points are shown on Figs. 13 and 14, respectively. A fit using Eq. (59) is shown on Fig. 13 and leads to $T_E(P_0) \simeq 550$ ns $> T_2$, which shows that part of the noise occurs at low frequency and is efficiently removed by the echo technique. Note that a naïve exponential fit of the bottom envelope of $p_E(\Delta t)$ would have given about the same $T_E$. Then, $T_E(P)$ values with a 30% uncertainty are extracted from each curve of Fig. 14 and reported on Fig. 15. A quantitative analysis of $T_E(P)$ is given below.

### D   Discussion of coherence times

A summary of all the coherence times ($T_2$, $T_E$) measured during free evolution using the various methods described above is given on Fig. 15. These results are in good agreement with each other and are comparable with those of our previous work[3]. As expected, $T_2$ is maximum at $P_0$ and decays by more than two orders of magnitude for $N_g$ or $\delta$ variations of 0.1 Cooper pairs or 0.3 phase turns, respectively. This result clearly validates the concept of the optimal working point. Moreover, while $T_2$ decreases rapidly when departing from $P_0$, the estimated sensitivity to $E_J$ noise given by Eq. (54) either decreases or stays constant, we thus conclude that $E_J$ noise has a negligible contribution to decoherence in this device at all working points except possibly at $P_0$. Figure 15 also shows that the improvement $T_E/T_2$ provided by the echo technique decreases from a factor of about 2 to about 1 when moving away from $P_0$ in the phase direction, and increases from about 2 to about 50 when moving in the charge direction. We try below to provide a quantitative understanding of these $T_2(P)$ and $T_E(P)$ variations, using simple model $S_\lambda(\omega)$ noise spectra for $\lambda = \delta, N_g$. Then, we discuss the decay of Ramsey fringes, $p_R(\Delta t)$, and of echo signals, $p_E(\Delta t)$, away from $P_0$. Finally, we discuss what might limit decoherence at $P_0$.

### 1   Noise spectral densities and $T_{2,E}(P)$ dependences

The fit to theory of the experimental $T_2(P)$ and $T_E(P)$ curves of Fig. 15 is performed in the following way. The dephasing factors $f_z$ are computed numerically according to the the-oretical expressions of section III and multiplied by the relaxation term $exp[-\Delta t/2T_1(P)]$, which is known from the independent measurements of Fig. 6 ; the coherence times correspond to a decay of these products by a factor $\exp(-1)$. First, we compute only the first order contribution of $\lambda$ noises (considered here as Gaussian) by numerical integration of Eqs. (17,19), using Eqs. (49, 51) for the $D_{\lambda,z}$'s. Microscopic charge and phase noises being characterized by $1/f$ spectra at low frequency and noises due to the driving and readout sub-circuits being characterized by white spectra below 10 MHz (see section II C), we start the fit using for $Sc_{Ng}(\omega)$ and $Sc_{\delta/2\pi}(\omega)$ linear combinations of $1/f$ and white spectral densities. Due to the divergence of the $1/f$ contributions as $\omega \to 0$, we introduce an infrared cutoff in the integration, $\omega_{ir} = 1/t_{meas}$, where $t_{meas} = 1$ s is the measurement time of a single data point in a Ramsey or echo signal. Note that although this cutoff could be defined more rigorously by taking into account the exact measuring protocols[3], this complication is of no benefit here because the computed coherence times depend only logarithmically on $\omega_{ir}$. At this stage, the fit (not shown) captures the $T_2(P)$ dependencies but does not capture the large gain $g = T_E/T_2$ observed far from $N_g = 1/2$. This problem was expected since the echo technique is inefficient in the presence of high frequency noise and because the gain deduced from Eqs. (21,22) in the case of a $1/f$ noise is $g \simeq \sqrt{\ln(t_{meas}/T_\varphi)/\ln(2)} \lesssim 5$ over the explored range of $T_2$. Consequently, $S_{Ng}(\omega)$ has to decrease *faster than* $1/f$ above a certain frequency. We thus introduce a high frequency sharp cutoff $\omega_c$ in the spectrum $Sc_{Ng}(\omega)$ as a new fitting parameter. The new fit (not shown) is then in fair agreement with the data except in the vicinity of $P_0$ where computed coherence times diverge due to the cancellation of the $D_{\lambda,z}$'s. Therefore, second order contributions have now to be included at this point using the $\partial^2\omega_{01}/\partial\lambda^2$'s given by Eqs. (50,52). For the sake of simplicity, $\lambda^2$ noises are first treated as Gaussian noises characterized only by their spectral densities $S_{\lambda^2}$ estimated from the auto-convolution of $S_\lambda$. This rough approximation leads to dephasing times at $P_0$ correct within a factor better than 2. We show in this way that the contribution of $\delta^2$ is completely negligible with respect to that of $N_g^2$. The calculation is then redone properly using

Eqs. (27, 31). Finally, the dephasing factors associated with $N_g$, $\delta$,  and $N_g^2$ are multiplied together. This procedure neglects the effect of correlations between $\lambda$ and $\lambda^2$, which are relevant only when both contributions are of same order, namely in a very narrow range in the vicinity of $P_0$. Moreover, our results are not affected by correlations between $N_g$ and $\delta$, which would exist if both noises were to be due to the same underlying mechanism, since the coupling coefficient $\partial^2\omega_{01}/\partial N_g\partial\delta$ for the cross noise $Sc_{Ng-\delta}(\omega)$ is zero along $(\delta, N_g = 1/2)$ and $(\delta = 0, N_g)$ lines. The final fit shown on Fig. 15 leads to $Sc_{Ng}(\omega) = 1.6 \ 10^{-6}/|\omega|$ for $|\omega| < \omega_c = 2\pi \times 0.4$ MHz and to $Sc_{\delta/2\pi}(\omega) = 0.9 \times 10^{-8}/|\omega| + 6 \times 10^{-16}/(\text{rad/s})$.

First we discuss the charge noise. The amplitude coefficient for the $1/f$ charge noise is in the range expected for a background charge noise $Sc_{Ng}^{BCN}$ of microscopic origin (see section II C). The high frequency cutoff $\omega_c$, necessary to provide even a qualitative fit, is an important result that had not been anticipated and that calls for a direct measurement of charge noise in the MHz range, perhaps using an rf-SET electrometer[44]. The white noise contribution to charge noise due to the gate impedance $Z_g$, deduced from Eqs. (5,16), provides a very large $T_\varphi \sim 300$ ms, this is compatible with our assumption of a high frequency cutoff. Note that this cutoff is only related to the *classical part* of the charge noise and does not preclude the possibility that charge TLFs might absorb energy at high frequencies, and thus relax the qubit[28].

We now turn to the phase noise. The presence of $1/f$ phase noise is similar to the unexplained flux noise found in SQUIDS (see section II C), although its amplitude corresponds here to a standard deviation $\sigma_{\Phi/\Phi0}$ about 10 times larger (spectral density 100 times larger) than that usually reported[30]. The value of the white phase noise of $\sim 6 \times 10^{-16}/(\text{rad/s})$ is about twice the estimated out-of-equilibrium noise expected from the AWG, whereas the impedance $Y_R$ is expected to contribute only a few percent more to this white spectrum. This white phase noise contribution is responsible for the low efficiency of echoes at $\delta \neq 0$, $N_g = 1/2$.

## 2 Temporal decays of Ramsey and echo signals

The phase and charge noises spectra mentioned above imply precise shapes for the temporal variations of Ramsey and echo signals. For $\delta \neq 0$, the dominant contribution to decoherence arises from the first order contribution of the phase noise $Sc_{\delta/2\pi}$. The numerical integration of Eqs. (17,19) predicts that the Ramsey function $f_{z,R}(\Delta t)$ involved in $p_R$ should be close to a Gaussian at small $|\delta|$ and should evolve towards an exponential at larger $|\delta|$, whereas the echo decay function $f_{z,E}(\Delta t)$ is expected to be almost exponential at all points. However, the contribution of the relaxation and of the second order noise at small $\delta$ on the first hand, and the contribution of the geometrical corrections included in Eqs. (58,59) on the second hand, favor exponential variations at short times $\Delta t < T_{2,E}$. Consequently, we find that the Ramsey signals are expected to decay more or less exponentially, as we observe on the left panels of Fig. 8, where the data were phenomenologically fitted by exponentially damped sinusoids. The echo variations shown on the left panels of Fig. 14 are exponential as expected, and are fitted accordingly. For $N_g \neq 1/2$, the dominant contribution to decoherence has been found to be a first order $1/f$ charge noise truncated at 0.4 MHz, which is actually 'quasistatic' according to section II C, since $\omega_c T_{2,E} << 1$. Consequently, if this noise is really Gaussian, $f_{z,R}$ should be given by Eq. (23), i.e. purely Gaussian. The decay should fit to Eq. (58), which includes the relaxation contribution and geometrical errors. Now, it was found that this equation does not fit the data well, even with unreasonably large geometrical errors, because oscillations survive much too strongly at large time $\Delta t > T_2$. Consequently, Fig. 8 shows an empirical fit with exponentials. This mismatch between the simple theory and the experiment might be attributed to the non Gaussian character of the $1/f$ charge noise, which is known to contain large discrete TLF's as already mentioned and as observed in the lineshapes. Depending on the distribution of these large fluctuators, Eq. (38) might be applicable. But such a formula gives an exponential decay for the ensemble average over all possible distributions of TLFs and is not supposed to describe quantitatively the non self averaging decay of single Ramsey samples (like those we have measured), for which a few TLFs are expected to dominate. Our experimental $p_R$'s could be compatible with a model which includes a dominant TLF inducing an initial Gaussian-like decay at small times $\Delta t < T_2$, and a large collection of further TLF's responsible for the exponential-like tail of the decay. In the same way, $f_{z,E}$ is expected to decay as $exp[-(\Delta t/T_E)^n]$ with $n \geq 3$ if the quasistatic $1/f$ noise is Gaussian. The rather exponential character of the measured $p_E$'s (see the right hand panels of Fig.14) also suggest that the non Gaussian character of the noise lowers the exponent $n$, as predicted by Eq. (39). On the other hand, the higher sensitivity of $p_E$ to geometrical errors (compared to $p_R$) also favors an exponential decay. To summarize, the decay times $T_{2,E}$ are well explained, but the temporal dependence of the functions $f_{z,E}(t)$ is not fully accounted for, possibly due to the non Gaussian character of the charge noise.

### 3  Decoherence at the optimal point $P_0$

Figure 15 shows that the best fit away from $P_0$ automatically leads to correct $T_{2,E}$ values at $P_0$. Knowing from the fitting procedure that the phase noise gives a negligible contribution to decoherence at this point, the following question arises: Can the quasistatic $1/f$ charge noise explain quantitatively the Ramsey decay shape at $P_0$? To answer this question, we plot on Fig. 7 the theoretical decay $exp\left[-\Delta t/2T_1\right] \left\{1 + \left[7.3\left(\Delta t/T_\varphi\right)\right]^2\right\}^{-1/4}$ where the second term is a simple rewriting of Eq. (27), with $T_\varphi = 620$ ns calculated from the fitted noise spectrum $S_{c_{Ng}}(\omega)$. This curve is seen to be in good agreement with the envelope of the best experimental $p_R(\Delta t)$ records. Whereas it is close to exponential at $\Delta t \lesssim T_2$, it predicts a significantly larger signal at long times, as we always observe. These results suggest that decoherence at the optimal working point $P_0$ is limited by second order microscopic static charge noise. Do the data in the frequency domain also support this conclusion? First, we observe on Fig. 11 that the resonance line at $P_0$ is asymmetric, which is a key feature of decoherence due to a second order noise at an optimal point. The line has indeed a tail on its higher frequency side because $N_g$ noise can only increase $\nu_{01}$, which is minimum at $P_0$.

More precisely, the intrinsic theoretical lineshape, i.e. the Fourier transform of Eq. (27), is non zero only at $\Delta\nu = \nu - \nu_{01} \geq 0$, is proportional to $\Delta\nu^{-1/2} \exp\left[-2\pi\Delta\nu T_\varphi/7.3\right]$ and is to be convolved with the Lorentzian lineshape due to relaxation. A subtle point already mentioned for $1/f$ noise is that decoherence data are actually dependent on the exact experimental protocol used to average them. In particular, $T_\varphi$ depends on the averaging time through the infrared cut-off introduced in the calculation of $\sigma_{N_g}$ [see Eq. (27)]. The 1Hz cut-off used for interpreting $p_R$ is no longer relevant for interpreting the lineshape, which was averaged over several records of 10 minutes each, with a precise tuning of $N_g$ before each record. The corresponding cut-off is of order of $1/(600\ \text{s})$ and the new $T_\varphi$ value analogous to the 620 ns used in the time domain is now 415 ns. Figure 11 shows the corresponding theoretical lineshape, which takes into account this $T_\varphi$ and $T_1$. This line is significantly narrower than the experimental one. This mismatch cannot be reduced by changing $T_\varphi$ (i.e. the infrared cut-off or the noise amplitude) since the line would be broadened only on its right side. Once more, this discrepancy might be attributed to the non Gaussian character of charge noise. To quantify the mismatch, we empirically fit the experimental line to the theoretical one convoluted with an additional Lorentzian. The width of this Lorentzian leading to the best fit corresponds to a characteristic decay time of 600 ns. This characteristic time can be used to place an upper bound for the $E_J$ noise. Indeed, attributing part of the additional contribution to this noise, assuming $Sc_{\delta E_J/E_J} = A/|\omega|$, and applying Eq. (17) with the same infrared cut-off as above, leads to $A < (3 \times 10^{-6})^2$, a value to be compared to the $(0.5 \times 10^{-6})^2$ mentioned in section (II C). In conclusion, decoherence at $P_0$ is dominated by microscopic charge noise at second order, the $E_J$ noise contributing at most for 40% and probably much less. Finally, we point out that pure dephasing is efficiently suppressed at $P_0$ with the echo technique, due to the ultra-violet cut-off of $Sc_{Ng}(\omega)$. Indeed, the measured $T_E = 550$ ns corresponds to a dephasing time $T_{\varphi,E} = 1.3\ \mu$s, partially hidden here by the short $T_1$ of the sample. A summary of these results is provided in Table I.

# V. DECOHERENCE DURING DRIVEN EVOLUTION

In the presence of a microwave driving voltage, the quantronium dynamics is best described in the rotating frame, as already mentioned in section III D. Due to decoherence, the precession of the effective spin is progressively dephased after a characteristic coherence time $\tilde{T}_2$ and, after some time $\tilde{T}_1$, the spin is almost depolarized because $\hbar\omega_R \ll kT$ in our experiment. In this section, we will describe the measurements of $\tilde{T}_2$ and $\tilde{T}_1$ at the optimal point $P_0$. We will compare them to the results of section III D and see if they can be understood from the noise spectra introduced in the preceding section.

## A   Coherence time $\tilde{T}_2$ determined from Rabi oscillations

The coherence time during driven evolution is directly obtained from the decay of Rabi oscillations since the ground state $|0\rangle = (|\tilde{0}\rangle + |\tilde{1}\rangle)/\sqrt{2}$ is a coherent superposition of the eigenstates under driven evolution. A series of Rabi experiments performed at the optimal point $P_0$ on resonance ($\Delta\omega = 0$) is shown in Fig. 16. These decays can be fit with exponentially damped sinusoids oscillating at $\omega_{R0}$, whose corresponding decay times $\tilde{T}_2$ are reported in Fig. 17 as a function of the Rabi frequency $\omega_{R0}/2\pi$, in the range $1 - 100$ MHz. The decay time $\tilde{T}_2$ is found to be almost constant at 480 ns under these conditions. This value being significantly shorter than $4T_1/3$, it gives access to $T_\nu = \Gamma_\nu^{-1} = 1.5 \pm 0.5$ $\mu$s using Eq. (45) $\tilde{\Gamma}_2 = 3\Gamma_1/4 + \Gamma_\nu/2$. Then, one deduces from $\Gamma_\nu \equiv \pi S c_{\delta\omega_z}(\omega_{R0})$ that $S c_{\delta\omega_z}(\omega)$ is, at $P_0$, constant at about $1.5 - 3 \times 10^5$ rad/s in the whole $1 - 100$ MHz range. Being obtained at the optimal point, the latter value should be explained either by the first order noise of $E_J$ or by second order noises $N_g^2$ and $\delta^2$. The $E_J$ noise, being of the $1/f$ type, cannot explain the constant $S c_{\delta\omega_z}(\omega)$. Then, assuming that the classical noise on $Ng$ is negligible at all frequencies above the low frequency cutoff of 0.4 MHz found in the previous section, the autoconvolution of $S c_{Ng}(\omega)$ has a negligible weight in the frequency range considered here and $S c_{\delta\omega_z}(\omega)$ can only be due the $\delta^2$ noise, whose spectral density is essentially given by the autoconvolution of the white $\delta$ noise introduced previously. Using a high frequency cutoff much higher than 100 MHz indeed leads to a constant $Sc_{\delta\omega_z}(\omega)$ as observed. Nevertheless, we have not found a plausible phase noise spectrum $Sc_{(\delta/2\pi)}(\omega)$ that could account for the measured value of $Sc_{\delta\omega_z}(\omega)$ using Eq. (50).

In order to test the $\tilde{\Gamma}_2(\eta)$ dependence predicted by Eq. (45), a series of Rabi precession experiments was also performed at $P_0$ as a function of the detuning $\Delta\omega/2\pi$, using a fixed microwave power corresponding to a Rabi frequency of $\omega_{R0}/2\pi = 15.4$ MHz on resonance. The data are also presented on Fig. 17 together with the $\tilde{\Gamma}_2$ expression given by Eq. (45), plotted using the $T_1$, $T_\varphi$, and $T_\nu$ values determined previously.

## B   Relaxation time $\tilde{T}_1$ determined from spin-locking experiments

The relaxation time $\tilde{T}_1$ can be obtained using the spin-locking technique developed in NMR. After having prepared the fictitious spin along an axis in the equatorial plane of the Bloch Sphere, the effective field is immediately oriented parallel or antiparallel to the spin. Experimentally, the spin is prepared along the $Y$ axis using a resonant ($\Delta\omega = 0$) $\pi/2$ pulse around the $X$ axis. A microwave gate voltage with a phase shifted by $\pm\pi/2$ is then applied so that the driving field is parallel (or antiparallel) to the prepared spin state, which becomes either $\left|\tilde{0}\right\rangle$ or $\left|\tilde{1}\right\rangle$, respectively. The polarization along the prepared direction then decays exponentially with a decay time $\tilde{T}_1$ called in NMR the relaxation time in the rotating frame[31]. A second $\pi/2$ or a $3\pi/2$ pulse is then applied around the $X$ axis after a variable delay in order to measure the remaining polarization in the rotating frame. This decay measured with a locking microwave field of $\omega_{R0}/2\pi = 24$ MHz is shown in Fig. 18, together with the envelope of a Ramsey signal measured at $\Delta\omega/2\pi = 8$ MHz and a relaxation signal recorded during free evolution. The evolution of the spin-locking signals towards equilibrium follows an exponential law with $\tilde{T}_1 \approx 550 \pm 50$ ns, irrespective of whether the spin is parallel or antiparallel to the locking field. This is because the energy splitting $\hbar\omega_{R0}$ of the levels $\left|\tilde{0}\right\rangle$ and $\left|\tilde{1}\right\rangle$ in the rotating frame is small, $\hbar\omega_{R0} \ll k_B T$. Using Eq. (42) $\tilde{\Gamma}_1 = \Gamma_\nu + \frac{1}{2}\Gamma_1$, one obtains again $T_\nu = 1.5 \pm 0.5 \ \mu$s, in agreement with the analysis of Rabi oscillations.

## VI. DECOHERENCE MECHANISMS IN THE QUANTRONIUM, PERSPECTIVES, AND CONCLUSIONS

### A  Summary of decoherence mechanisms in the quantronium

We have characterized decoherence in a superconducting qubit circuit, the quantronium, using techniques adapted from NMR. We have presented a general framework that describes these experiments. As expected, we have found that quantum coherence of the quantronium is minimum at the so-called optimal point $P_0$, where the decay laws of the transverse polarization can be significantly non-exponential, particularly in the presence of $1/f$ noise. Similar and complementary analyses of decoherence have now been performed in other Josephson qubits[28,45–47]. We have also derived the noise spectra that characterize the sources leading to decoherence of the quantronium, at and away from $P_0$. We have shown that coherence is mainly limited by dephasing due to charge and phase noises of microscopic origin; and that relaxation also contributes. An important feature of our analysis is the introduction of a *high-frequency cutoff* at about 0.5 MHz for the classical part of the *charge noise* spectrum. Finally, it was shown that in our qubit with $E_J \sim E_C$, second order charge noise is dominant at $P_0$.

Although our semi-empirical approach obviously did not aim at providing any definite clues about the exact nature of the microscopic defects responsible for the noise spectra invoked to explain decoherence, the subject is very important and deserves further studies. To improve our understanding, more refined models could be built including a finite set of strongly coupled slow TLFs, with a close-to-continuous background of weakly coupled ones, including the non-Gaussian nature of their noise (see[39,43]).

Finally, we point out that some of the NMR methods that we have used to characterize decoherence in our circuit provide tools for improving coherence in a qubit. We now discuss the interest of maintaining quantum coherence with these methods, and how far we are from meeting the requirements for elementary quantum computing.

## B   Does driving the qubit enhance its coherence?

The observation that $\tilde{T}_2 > T_2$ suggests that the coherence is improved by driving the qubit. But what are the reason and the meaning of this observation? The gain is actually due to the divergence of the noise spectral density $Sc_{\delta\omega}(\omega)$ at low frequency. Indeed, when the Rabi frequency is large enough, the low frequency fluctuations $\delta\omega$ are not effective because the eigenstates $\left|\tilde{0}\right\rangle$ and $\left|\tilde{1}\right\rangle$ follow adiabatically the fluctuations of the effective driving field, as predicted by Eq. (42). Consider now that a coherent superposition of the two eigenstates in the rotating frame, $\alpha\left|\tilde{0}\right\rangle + \beta\left|\tilde{1}\right\rangle$, has been prepared and that a Rabi field is applied. The superposition then evolves at the Rabi frequency, and the initial state is retrieved periodically with a coherence time $\tilde{T}_2 > T_2$. By encoding the qubit in the basis $\left(\left|\tilde{0}\right\rangle, \left|\tilde{1}\right\rangle\right)$, quantum coherence is thus maintained during a longer time than for free evolution. Rabi precession provides a direct test of this result because the ground state is an equal weight superposition $|0\rangle = \left|\tilde{0}\right\rangle + \left|\tilde{1}\right\rangle$. When a coherent superposition of these eigenstates is prepared, and a locking field applied afterwards, the initial state is frozen with coherence time $\tilde{T}_2 > T_2$ and mixing time $\tilde{T}_1 > T_1$. Although these two examples show that a qubit state can indeed be stored during a longer time by driving it, it is clear that the qubit cannot be used at will during its driven evolution.

The echo technique can also be regarded as a 'soft' driving of the qubit aiming at reducing decoherence. As shown in section IV C, it indeed removes the effect of the low frequency fluctuations of $\delta\omega$. It can be figured as a time-reversal operation that compensates frequency changes that are almost static over the duration of the pulse sequence. This method is in fact more general, and the repeated application of $\pi$ pulses can compensate for frequency fluctuations over longer durations. This so-called bang-bang technique in NMR could be used for qubits provided that the coherence loss due to the pulses is small enough[37].

## C   Coherence and quantum computing

Although the simple methods mentioned above could help in reducing qubit decoherence, real quantum error correcting codes are mandatory for quantum computing. These codes are known to require error rates smaller than about $10^{-4}$ depending on the nature of the errors for each logic gate. Presently, the gate error rate can be estimated at a few % for single qubit gates (e.g. the quantronium), and significantly more for two qubit gates such as coupled Cooper pair boxes[48]. The coherence time is about a few hundreds times longer than the duration of a single qubit gate operation in the quantronium, and would be at best a few tens times the duration of a two qubit gate. Since decoherence is equivalent to making errors, the quantronium requires an improvement of coherence time by two or three orders of magnitude. The operation of a quantum processor based on this qubit circuit, or on any other one presently developed, thus appears to be a significant challenge.

This is, however, not a reason to give up because conceptual and technical breakthroughs are to be expected in the rather new field of quantum circuits. Progress in junction fabrication might in particular lead to a significant increase of coherence times in Josephson qubit circuits. Furthermore, it is already close to possible to run simple algorithms such as Grover's search algorithm, and to address important questions in quantum mechanics. The extension of quantum entanglement from the microscopic to the macroscopic world, and the location and nature of the frontier between the quantum and classical worlds, are two essential issues. For instance, the accurate measurement of the correlations between two coupled qubits in order to test the violation of Bell's inequalities could indeed probe whether or not the collective variables of qubit circuits follow quantum mechanics. Such an experiment will become possible as soon as a high fidelity readout is available, which is clearly an important step to pass.

## ACKNOWLEDGMENTS

We acknowledge numerous discussions in the Quantronics group, discussions with G. Falci, the technical help of P.F. Orfila, P. Sénat and J.C. Tack, the support of the European project SQUBIT, of Yale University (grant DAAD 19-02-1-0044), of the Landesstiftung BW and of the Dynasty foundation. P.J. Meeson acknowledges a Marie Curie fellowship.

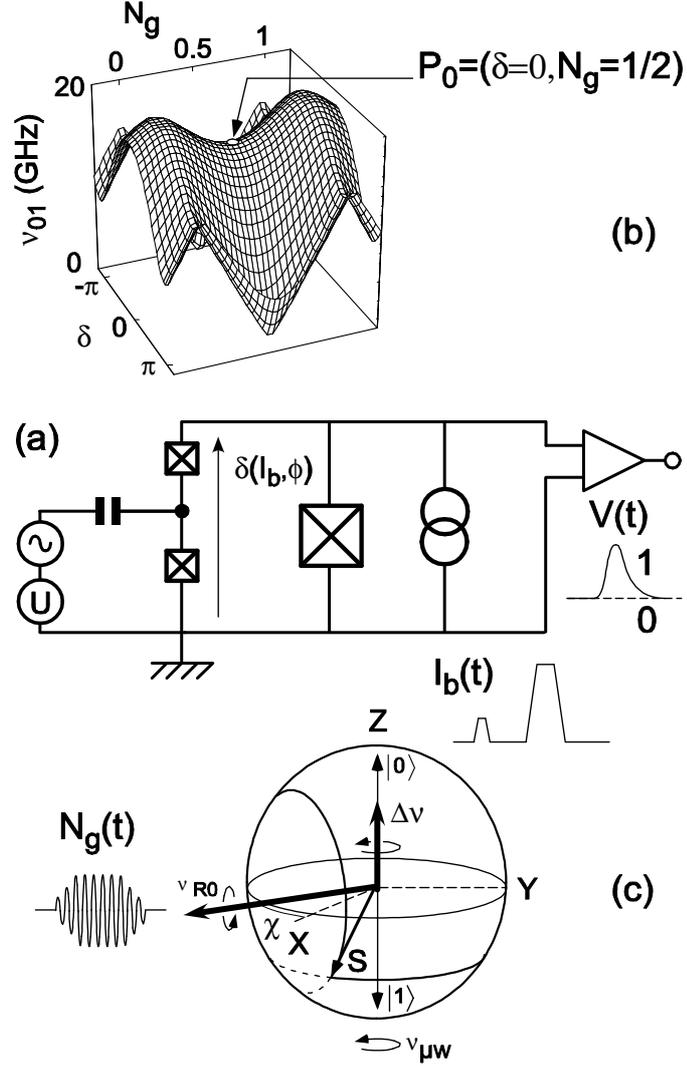

FIG. 1: (a): circuit diagram of the quantronium. The Hamiltonian of this circuit is controlled by the gate-charge $N_g \propto U$ on the island between the two small Josephson junctions and by the phase $\delta$ across their series combination. This phase is determined by the flux $\phi$ imposed through the loop by an external coil, and by the bias-current $I_b$. The two lowest energy eigenstates form a quantum bit whose state is read out by inducing the switching of the larger readout junction to a finite voltage V with a bias-current pulse $I_b(t)$ approaching its critical current. (b): qubit transition frequency $\nu_{01}$ as a function of $\delta$ and $N_g$. The saddle point $P_0$ indicated by the arrow is optimal for a coherent manipulation of the qubit. (c): Bloch sphere representation in the rotating frame. The quantum state is manipulated by applying resonant microwave gate pulses $N_g(t)$ with frequency $\nu_{\mu w}$ and phase $\chi$, and/or adiabatic trapezoidal $N_g$ or $I_b(t)$ pulses. Microwave pulses induce a rotation of the effective spin $S$ representing the qubit around an axis in the equatorial plane making an angle $\chi$ with $X$, whereas adiabatic pulses induce rotations around $Z$.

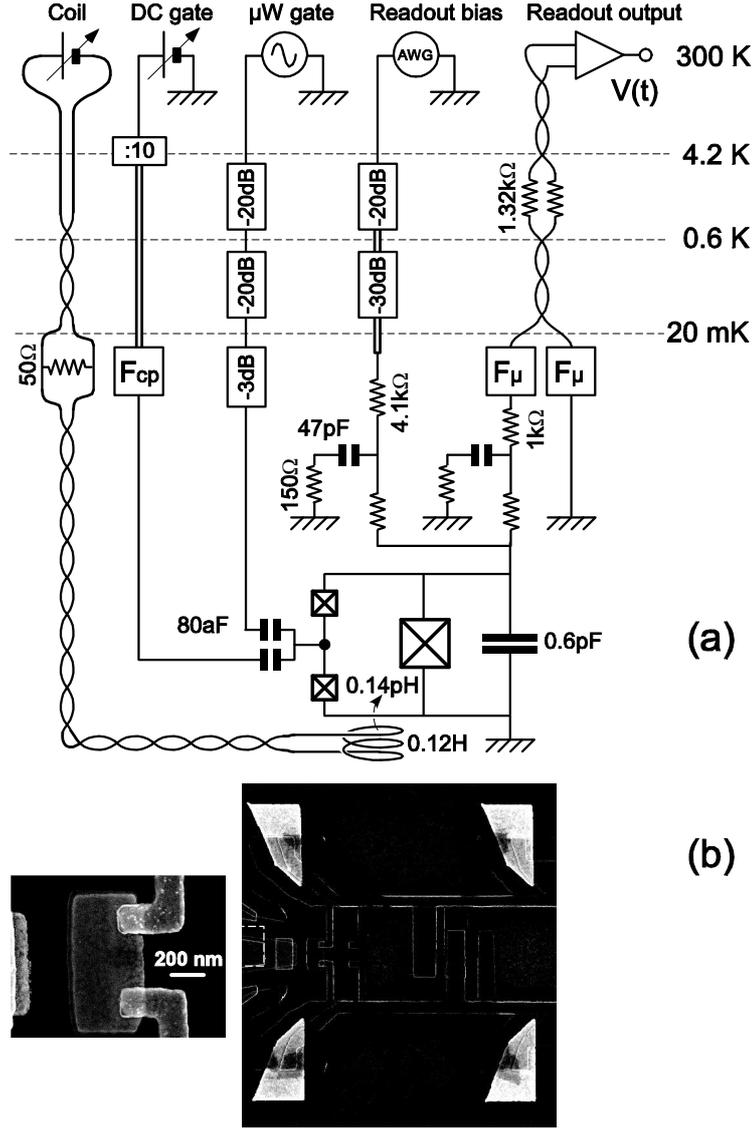

FIG. 2: (a): Schematics of the experimental setup used in this work with temperatures indicated on the left. Rectangles labelled in dB are 50 Ω attenuators whereas rectangle labelled ":10" is a high impedance voltage divider by 10. Squares labelled Fcp and Fμ are copper powder filters and microfabricated distributed RC filters, respectively. Single lines, double lines and twisted pairs are 50 Ω coaxes, lossy coaxes made of a manganin wire in a stainless steel tube, and shielded lossy manganin twisted pairs, respectively. (b): Scanning Electron Microscope (SEM) pictures of the sample. The whole aluminum loop (left) of about 5.6 $\mu$m$^2$ is defined by 200 nm wide lines and includes a 890 nm × 410 nm island (right) delimited by two 160 nm × 160 nm junctions, and a 1.6 $\mu$m × 500 nm readout junction. Note also the presence of gold quasiparticle traps (bright pads).

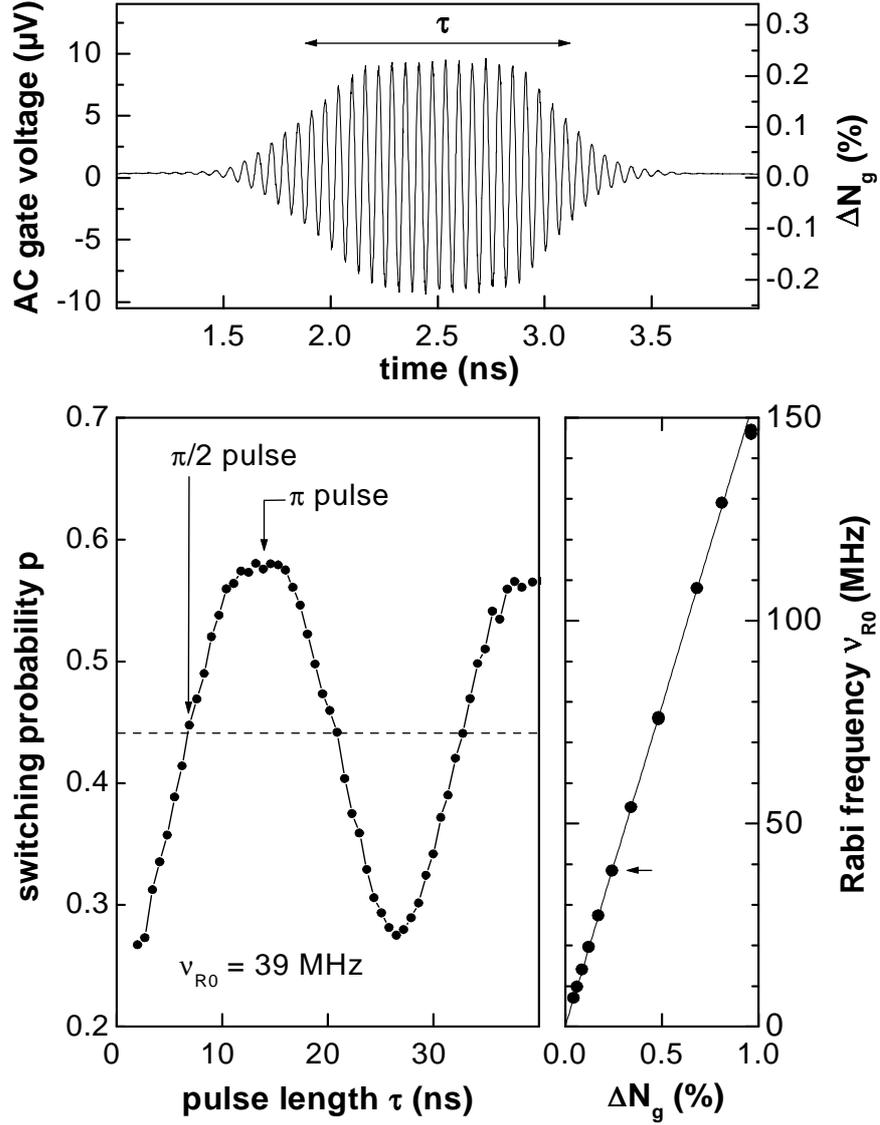

FIG. 3: Top: oscillogram of a typical microwave gate pulse, measured at the top of the cryostat. The arrow indicates the effective duration $\tau$ of the pulse. Bottom: the Rabi precession of the qubit state during a microwave pulse results in oscillations of the switching probability $p$ with the pulse length $\tau$, at a frequency $\nu_{R0} = \omega_{R0}/2\pi$ proportional to the reduced microwave amplitude $\Delta N_g$ (right). The two arrows in the left panel correspond to the so-called $\pi$ and $\pi/2$ pulses used throughout this work. The arrow in the right panel indicates the point that corresponds to the data shown in the left panel.

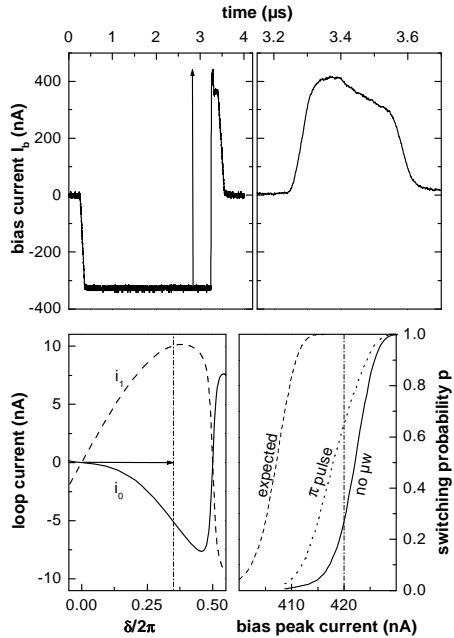

FIG. 4: Readout of the quantronium. Top left: full $I_b(t)$ variation measured at the top of the cryostat, when the qubit is operated at the optimal point (see text) and read out with maximum sensitivity. The current is first kept at zero to avoid heating in the bias resistor, then pre-biased at a negative value that corresponds to $\delta = 0$, then increased in about 50 ns to a value close to the critical current of the readout junction, and maintained at this value during about 100 ns, a time period over which the switching of the junction may occur. $I_b$ is then slightly lowered and maintained at this lower value to let the voltage develop along the measuring line if the junction has switched. Finally it is set back to zero. Top right: detail of readout pulse (but without negative pre-bias), measured at room temperature at the bottom of the bias line, before cooling the cryostat. Bottom left: persistent currents in the loop for the $|0\rangle$ (solid line) and $|1\rangle$ (dashed line) states, computed at $N_g = 0$ as a function of $\delta$, using the measured sample parameters. The vertical dotted-dashed line indicates the readout point $\delta_M$ where the experimental difference $i_1 - i_0$ was found to be maximum. Bottom right: variation of the switching probability $p$ with respect to the peak current, measured without microwave (solid line), measured after a $\pi$ microwave pulse (dotted line), and calculated from the sample parameters for state $|1\rangle$ (dashed line). The vertical dotted-dashed line indicates a maximum fidelity of 0.4 ( instead of the expected 0.95), obtained with the pulse shown in the top panels. The two arrows of the upper and lower left panels indicate the adiabatic displacement in $\delta$ between operation and readout of the qubit.

FIG. 5: Schematic drawing of the noise sources responsible for decoherence in the quantronium. These sources are either coupled to $E_J$, $N_g$ or $\delta$. In part they are of microscopic nature like the two-level fluctuators (TLF) inside the junction that induce Ej variations, like charged TLF (represented as a minus sign in a small double arrow) coupled to $N_g$, or like moving vortices ($\Phi_{micro}$) in the vicinity of the loop. The macroscopic part of the decoherence sources is the circuitry, which is represented here as an equivalent circuit as seen from the qubit. The relevant resistances and temperature of the dissipative elements are indicated. Capacitance with no label represent a shunt at the qubit frequency and an open circuit at frequencies below 200 MHz.

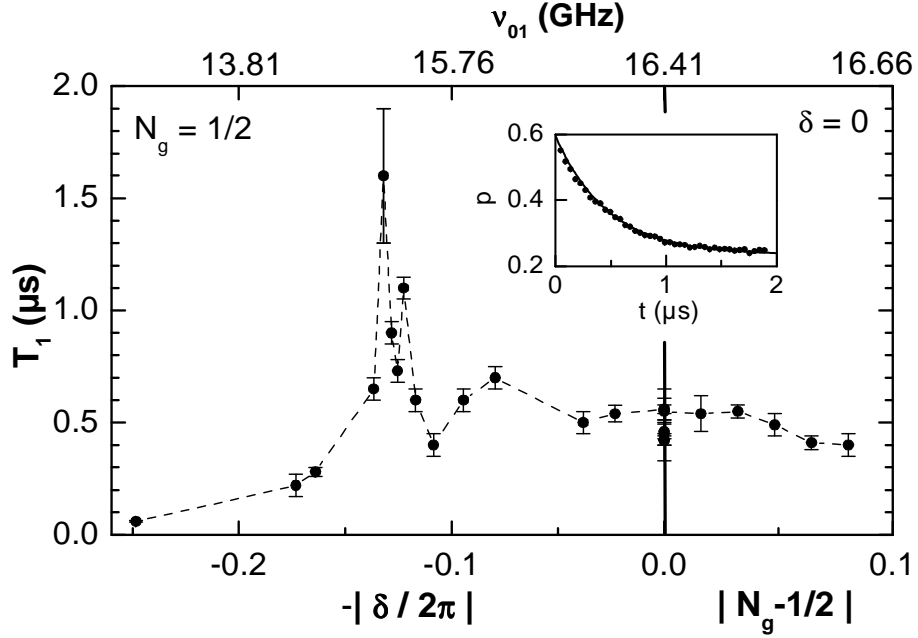

FIG. 6: Experimental $T_1$ values measured at $N_g = 1/2$ as a function of $\delta$ (left panel), and at $\delta = 0$ as a function of $N_g$ (right panel). The vertical line separating the two panels corresponds to the optimal point $P_0 = (N_g = 1/2, \delta = 0)$. The dashed line joining the points is a guide for the eye. The correspondence between $\delta$, $N_g$ and $\nu_{01}$ is given by the upper horizontal axis. Inset: Example of $T_1$ measurement. The switching probability $p$ (dots) is measured as a a function of the delay $t$ between a $\pi$ pulse and the readout pulse. The fit by an exponential (full line) leads to $T_1$ (0.5 $\mu$s at $P_0$ in this example).

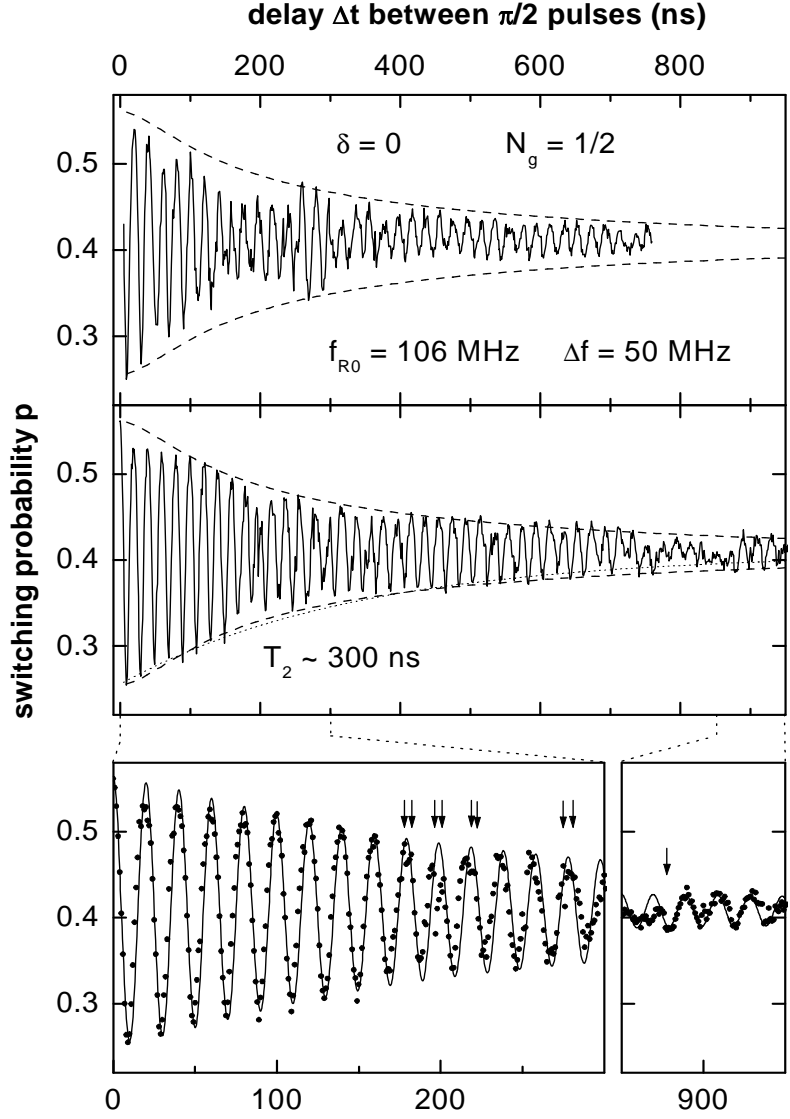

FIG. 7: Ramsey signals at the optimal point $P_0$ for $\omega_{R0}/2\pi = 106$ MHz and $\Delta\nu \approx 50$ MHz, as a function of the delay $\Delta t$ between the two $\pi/2$ pulses. Top and middle panels: solid lines are two successive records showing the partial irreproducibility of the experiment. Dashed lines are a fit of the envelope of the oscillations in the middle panel (see text) leading to $T_2 = 300$ ns. The dotted line shows for comparison an exponential decay with the same $T_2$. Bottom panels: zoom windows of the middle panel. The dots represent now the experimental points whereas the solid line is a fit of the whole oscillation with $\Delta\omega/2\pi = 50.8$ MHz. Arrows point out a few sudden jumps of the phase and amplitude of the oscillation, attributed to strongly coupled charged TLFs.

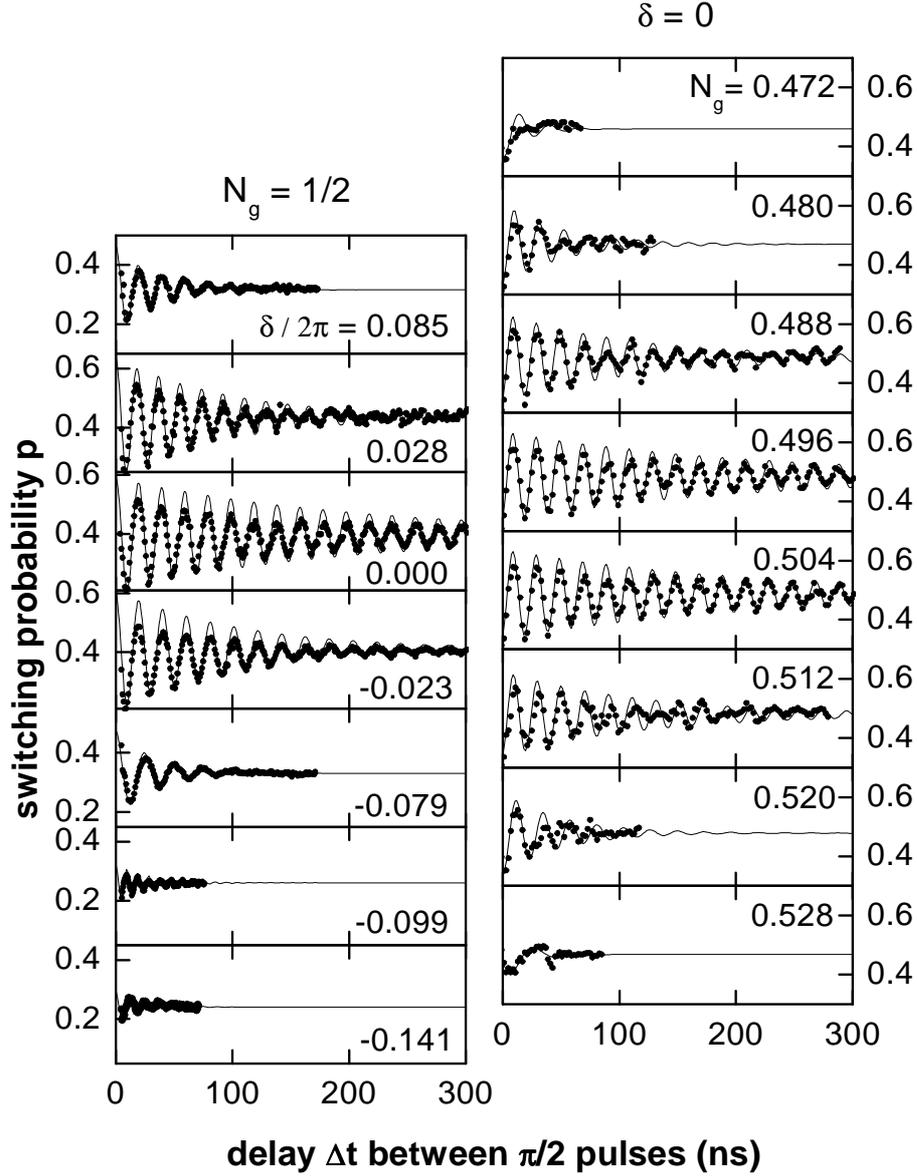

FIG. 8: Ramsey oscillations as a function of the delay $\Delta t$ between the two $\pi/2$ pulses, for different working points located on the lines $N_g = 1/2$ (left column) and $\delta = 0$ (right column). The Rabi frequency is $\omega_{R0}/2\pi = 162$ MHz for all curves. The nominal detunings $\Delta\nu$ are 50, 53, 50, 50, 40, 100 and 80 MHz (left, top to bottom) and $35 - 39$ MHz (right). Dots are experimental points whereas full lines are exponentially damped sinusoids fitting the experimental results and leading to the $T_2$ values reported on Fig. 15.

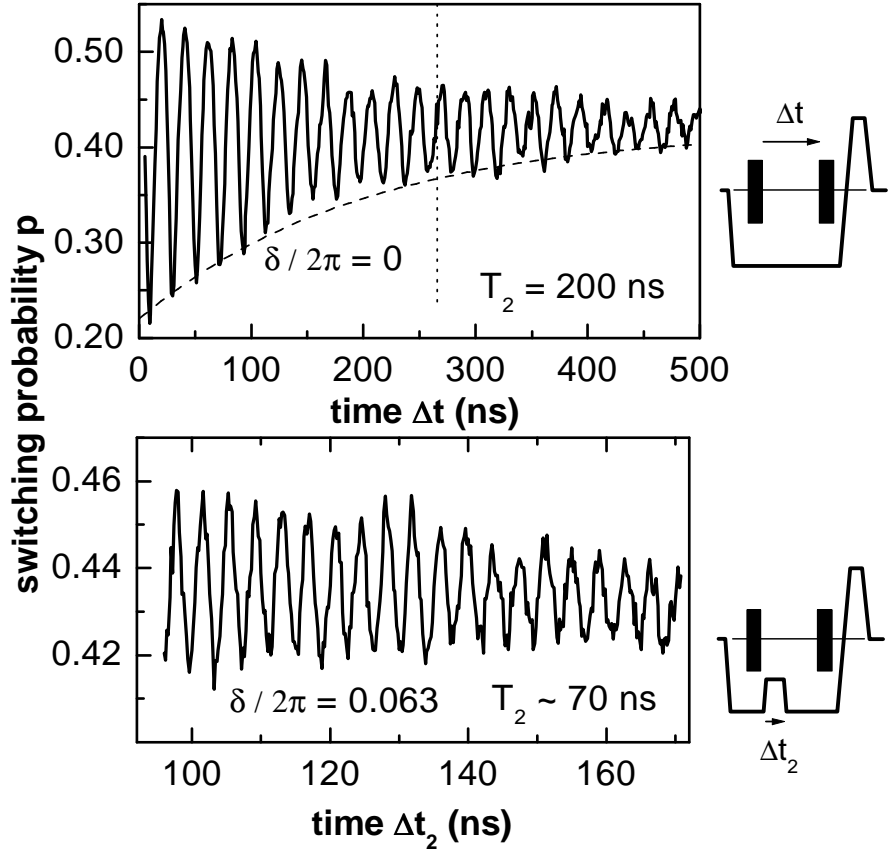

FIG. 9: Phase "detuning pulse technique" for measuring $T_2$. Top: Ramsey signal at the optimal point $P_0$, with $\Delta\nu \simeq 50$ MHz, when no detuning dc pulse is applied. The dashed line corresponds to an exponential decay with $T_2(P_0) = 200$ ns. Bottom: signal obtained with a delay $\Delta t = 275$ ns between the two $\pi/2$ pulses (corresponding to the dashed vertical line of the upper panel) and with an adiabatic current pulse maintaining $\delta/2\pi = 0.063$ during a time $\Delta t_2$. The oscillation of the signal with $\Delta t_2$ decays with a characteristic time of about 70 ns (note the different horizontal scales on the two graphs). The pictograms on the right illustrate the two $\pi/2$ microwave pulses and the $I_b(t)$ signal.

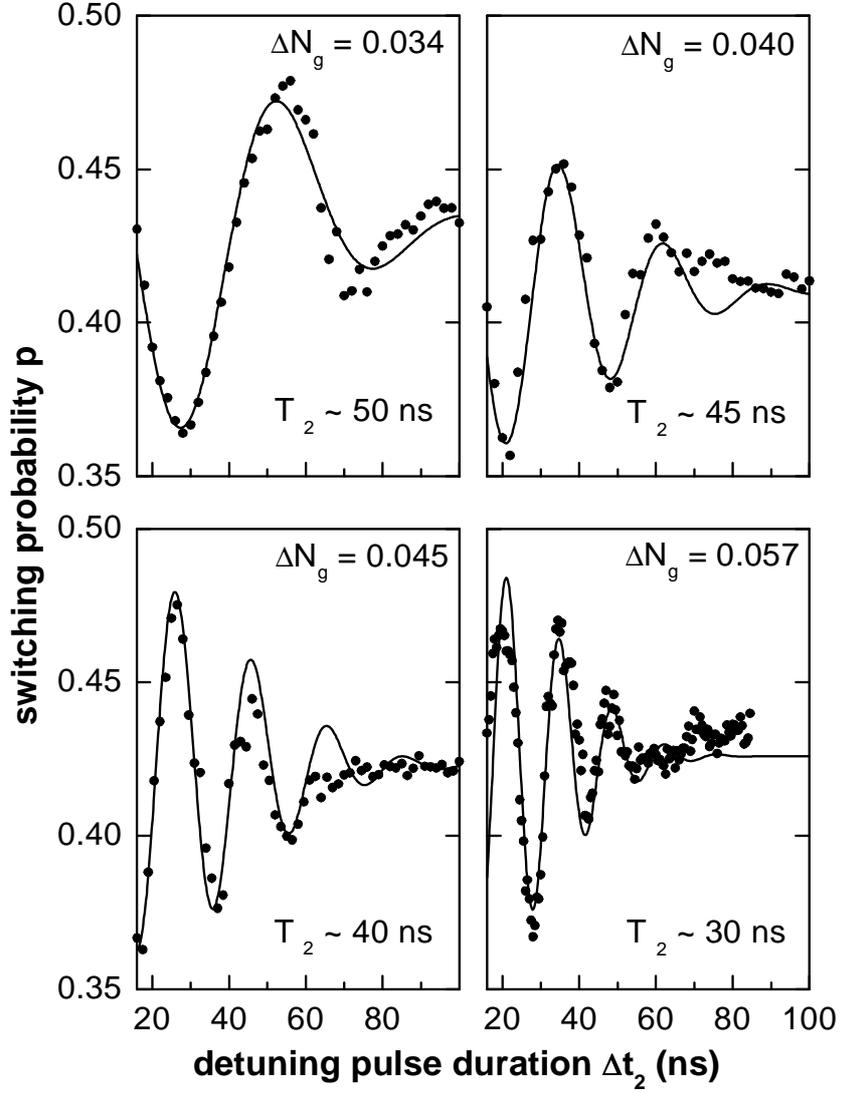

FIG. 10: Charge "detuning pulse technique" used for measuring $T_2$ at four points $P = (0, 1/2 + \Delta N_g)$. Dots are experimental points whereas full lines are fits using "Gaussian damped" sinusoids at frequencies $\Delta\omega(\Delta N_g)$. The extracted $T_2$ are indicated on each panel.

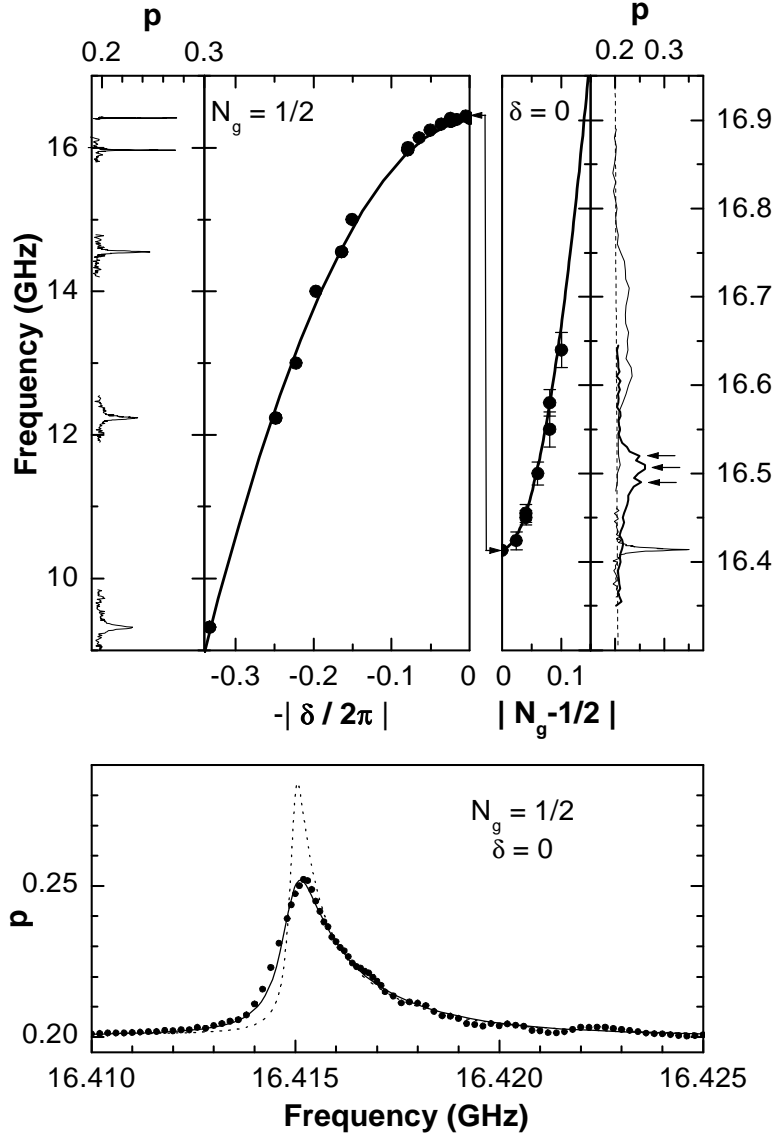

FIG. 11: Top panels: lineshape (thin lines) and central position (dots) of the resonance lines as a function of $\delta$ at $N_g = 1/2$ (left) and as a function of $N_g$ at $\delta = 0$ (right). The optimal point $P_0$ corresponds to the double-arrow in the center of the graph. Note the two different vertical scales and the occasional sub-structure of resonance lines pointed out by small arrows. Bold lines are fits of the peak positions leading to $E_J = 0.87\ k_B$ K, $E_C = 0.66$ $k_B$ K and $d < 13\%$. Bottom panel: asymmetric lineshape recorded (dots) at $P_0$ with a microwave power small enough to desaturate the line. The dashed line is the theory, with a $T\varphi$ that corresponds to that of Fig. 7. The solid line is the convolution of this theoretical line and of a Lorentzian corresponding to a decay time of 600 ns (see text).

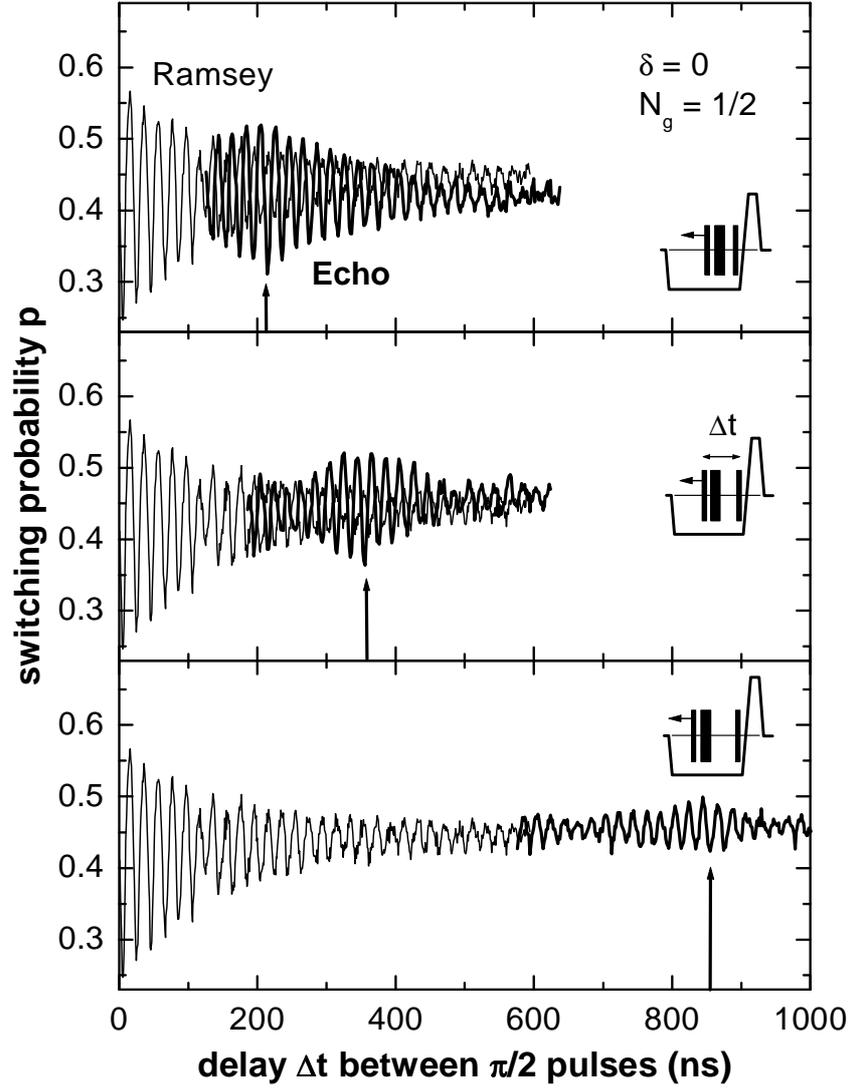

FIG. 12: Spin echoes (bold lines) obtained at the optimal point $P_0$ with a detuning $\Delta\nu \simeq 50$ MHz. Pictograms illustrate the experimental protocol: the delay $\Delta t_3$ between the $\pi$ pulse and the last $\pi/2$ pulse is kept constant while altering the timing of the first $\pi/2$ pulse. Each panel corresponds to a different $\Delta t_3$. Vertical arrows indicate the sequence duration $\Delta t = 2\Delta t_3$ for which the echo amplitude is expected to be maximal and where $p = p_E$ is minimum. For a sake of comparison, the corresponding Ramsey signal (thin lines) is shown in all panels.

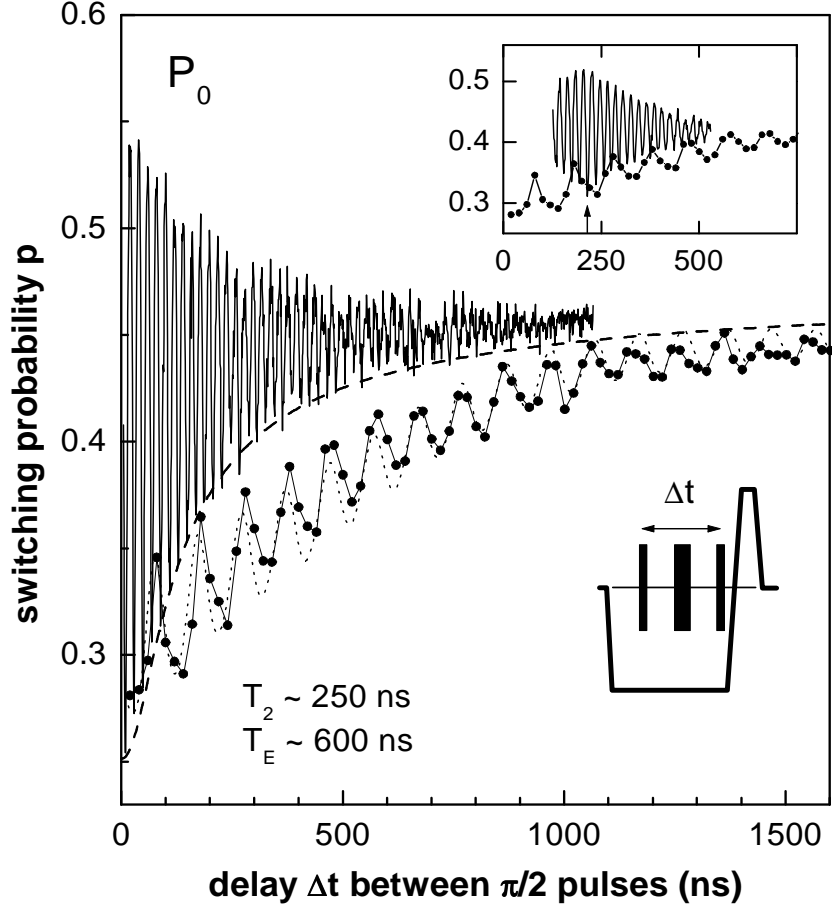

FIG. 13: Echo signal $p_E$ (linked big dots) measured at the optimal point $P_0$ by keeping a $\pi$ pulse precisely in the middle of the sequence while sweeping the sequence duration $\Delta t$ (pictogram). The Rabi frequency is $\omega_{R0}/2\pi = 130$ MHz and the detuning $\Delta\nu = 20$ MHz. For comparison, the Ramsey signal (oscillating line) and its envelope (dashed line leading to $T_\varphi = 450$ ns) are also shown. The dotted line is a fit of $p_E$ that leads to the characteristic decay time of $f_{z,E}$, $1.2\mu$s, and that shows that the $\pi/2$ pulses were actually 15% too short whereas the $\pi$ pulse was correct. The resulting echo time is $T_E \sim 600$ ns. Inset: comparison between $p_E$ (linked dots) and the echo signal recorded with a fixed $\pi$ pulse (solid line), as presented in Fig. 12).

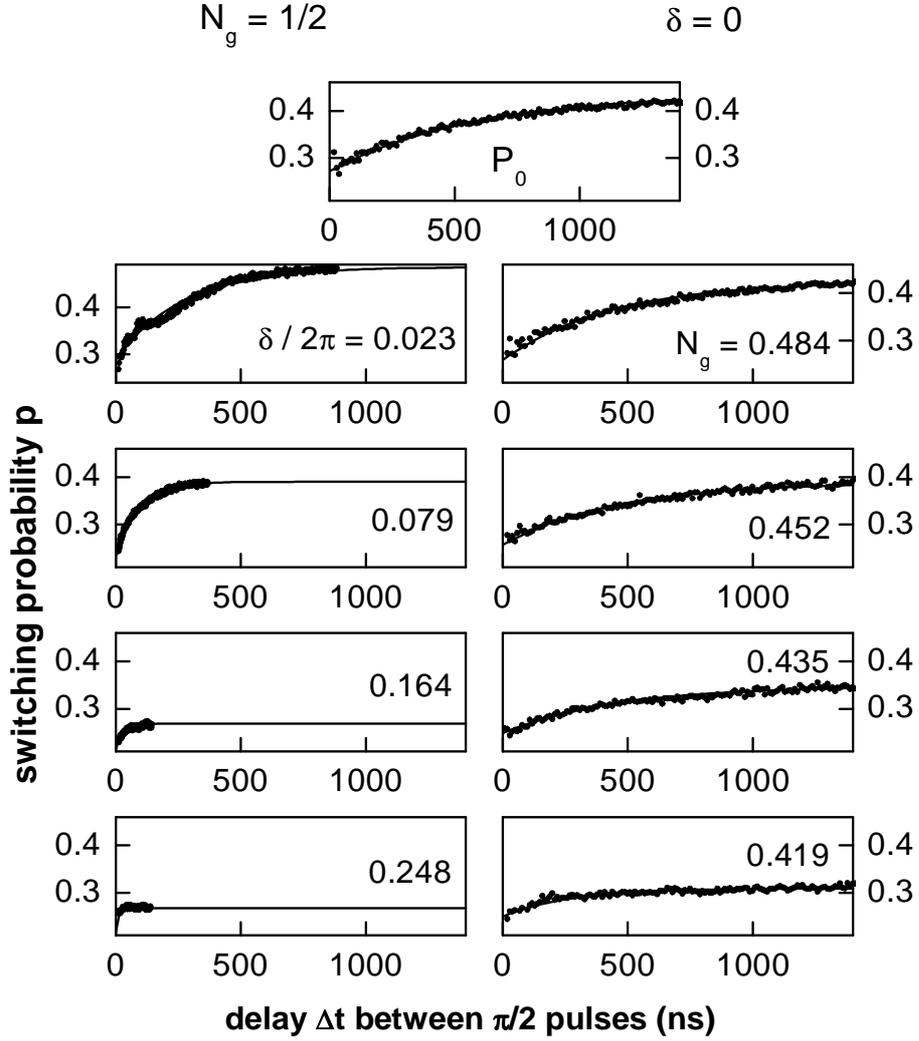

FIG. 14: Echo signals $p_E(\Delta t)$ measured (dots) at different working points indicated in each panel, with a Rabi frequency $\omega_{R0}/2\pi = 140$ MHz and $\Delta\nu \approx 50$ MHz. Full lines are exponential fits leading to $T_E$ values reported on Fig. 15. Note that the amplitude of the signal depends on the working point.

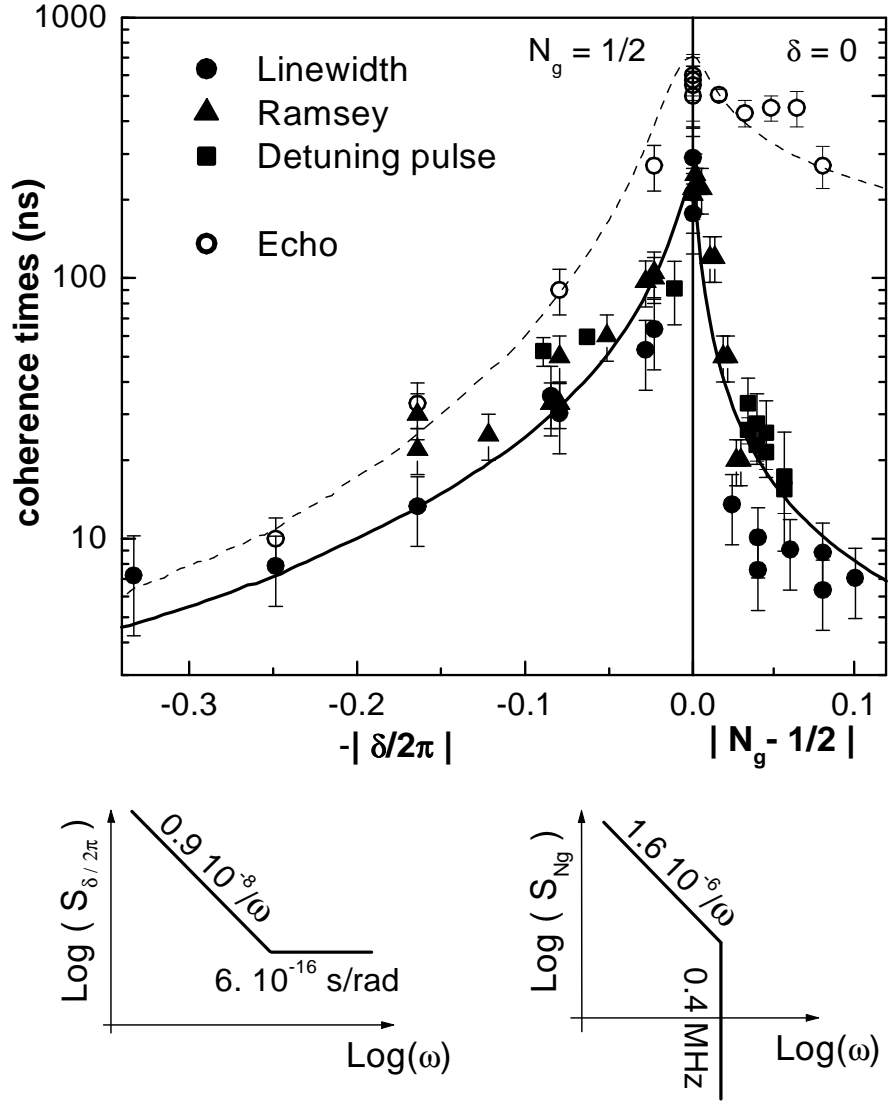

FIG. 15: Echo times $T_E$ (open circles) and coherence times $T_2$ measured from the resonance linewidth (solid dots), from the decay of Ramsey signals (triangles), and from the detuning pulse method (squares), at $N_g = 1/2$ as a function of $\delta$ (left panel) and at $\delta = 0$ as a function of $N_g$ (right panel). The full and dashed lines are best fits (see text) of $T_E$ and $T_2$ times, respectively, leading to the phase and charge noise spectral densities depicted at the bottom. The spectra are even functions of $\omega$.

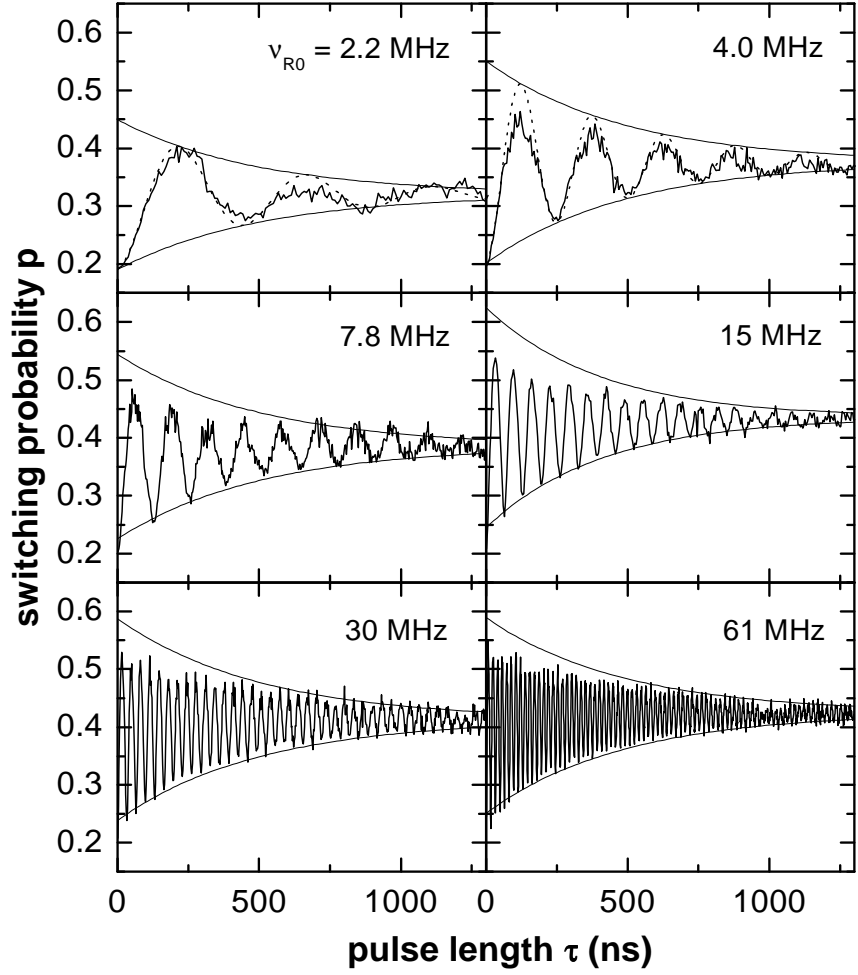

FIG. 16: Decay of the Rabi signals at the optimal point $P_0$ for different Rabi frequencies $\nu_{R0} = \omega_{R0}/2\pi$ (i.e. different microwave powers). The full lines are exponential fits of the lower envelopes leading to the $\widetilde{T}_2$ values reported in Fig. 17.

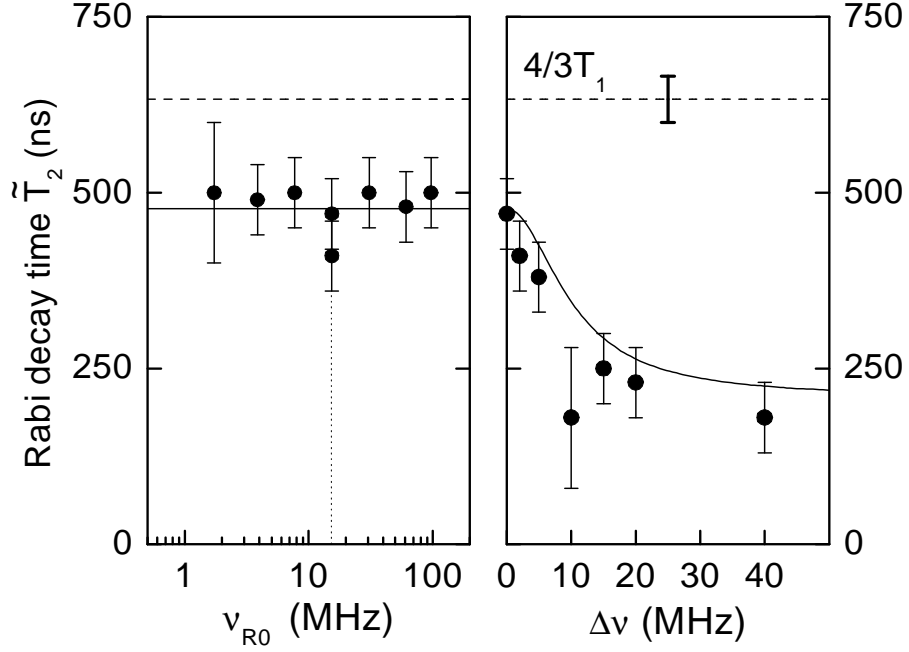

FIG. 17: Characteristic decay times $\widetilde{T}_2$ of the Rabi oscillations at the optimal point $P_0$, as a function of the Rabi frequency $\nu_{R0}$ (left panel) at zero detuning $\Delta\nu$, and as a function of $\Delta\nu$ (right panel) at $\nu_{R0} = 15.4$ MHz (dotted vertical line). $\widetilde{T}_2(\nu_{R0}, \Delta\nu = 0)$ turns out to be a constant of order 0.48 $\mu$s (left solid line). The difference with $4/3T_1$ leads to an estimate for $T_\nu = 1/\Gamma_\nu$. The right solid line corresponds to Eq. 45 plotted using the experimentally determined values of $T_\varphi$, $T_1$, and $T_\nu$.

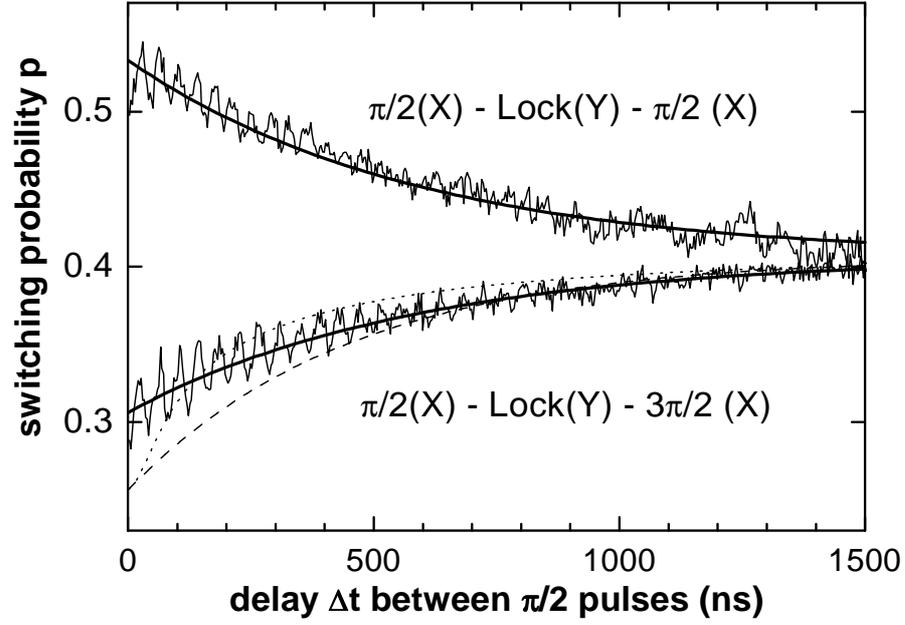

FIG. 18: Spin locking signals (oscillatory lines) obtained at the optimal point $P_0$, using a detuning $\Delta\nu = 8$ MHz, a locking microwave power corresponding to 24 MHz, and a final microwave pulse of $\pi/2$ (top) or $3\pi/2$ (bottom). The bold solid lines are exponential fits corresponding to $\widetilde{T_1} \backsim 580$ ns. For comparison, the Ramsey envelope (dotted line with $T_2 \backsim 250$ ns) and the longitudinal relaxation (dashed line with $T_1 \backsim 450$ $ns$) are shown.

| | $N_g$ noise | | $\delta$ noise | | $E_J$ noise | Total | Measured | |
|---|---|---|---|---|---|---|---|---|
| | Gate circuit | TLFs | Readout circuit | Micro | Micro | | | |
| $S_\lambda(\omega_{01})$ (s/rad) | $(1 - 3\ 10^{-9})^2$ | ? | $(\,80 - 20\ 10^{-9})^2$ | ? | ? | | | |
| $D_{\lambda,\perp}$ $(10^{11}$ rad/s) | 1.93 | | $3.8d \simeq 0.12$ | | $0.54\,\lvert N_g - \frac{1}{2}\rvert$ | $\leq$ 1.5 - 5 $\mu$s | 0.5 - 1 $\mu$s | |
| $T_1$ | $15 - 2\ \mu$s | ? | $3 - 6\ \mu$s | ? | $\infty$ ? | | | |
| $Sc_\lambda(\lvert\omega\rvert)$ (s/rad) | $(30\ 10^{-9})^2$ | $(1.3\ 10^{-3})^2/\lvert\omega\rvert$ $\lvert\omega/2\pi\rvert < 0.4$ MHz | AWG $\lessapprox (30\ 10^{-9})^2$ | $(1\ 10^{-4})^2/\lvert\omega\rvert$ | $< (3\ 10^{-6})^2/\lvert\omega\rvert$ | | | |
| Sensitivity $D_{\lambda,z}$ $(10^{11}$ rad/s) | 0 | | 0 | | 0.85 | | | |
| $\partial^2\omega_{01}/\partial\lambda^2$ | 2.9 | | $-8.5$ | | - | | | |
| $T_\varphi\,(P_0)$ | $\sim 300$ ms | 0.6 $\mu$s | $\sim 7\ \mu$s | $\sim 50\ \mu$s | $> 0.7\ \mu$s | $\lessapprox 0.6\ \mu$s | 0.6 $\mu$s | $T_2 = 0.3\ \mu$s |
| $T_{\varphi,E}\,(P_0)$ | irrelevant | 1.3 $\mu$s | $\sim 7\ \mu$s | irrelevant | ? | 1.3 $\mu$s | 1.3 $\mu$s | $T_E = 0.55\ \mu$s |

TABLE I: Summary of the relevant spectral densities and characteristic times characterizing decoherence at the optimal point $P_0$. Units are indicated in the left column for angular frequencies $\omega$ expressed in rad/s.